\documentclass[pra,twocolumn,superscriptaddress,showpacs,floatfix]{revtex4}

\usepackage{graphicx}  
\usepackage{amsmath,amssymb}
\newcommand{\ds}{\displaystyle}
\newcommand{\ket}[1]{|#1\rangle}
\newcommand{\bra}[1]{\langle#1|}
\newcommand{\tr}[2][]{\mathalpha{\mathrm{tr}}^{\ }_{#1}\left\{#2\right\}}
\newcommand{\expect}[1]{{\left\langle#1\right\rangle}}
\newcommand{\muB}{\mu_{\textsc{b}}}
\newcommand{\IP}[1]{\widetilde{#1}}
\newcommand{\bvec}[1]{\boldsymbol{#1}}
\newcommand{\unitdyad}{\mathbf{1}}
\newcommand{\erf}{{\mathalpha{\mathrm{erf}}}}

\begin{document}

\title{Long-lived qubit from three spin-1/2 atoms}

\author{\surname{Han} Rui}
\affiliation{Centre for Quantum Technologies, %
National University of Singapore, 3 Science Drive 2, %
Singapore 117543, Singapore}

\author{Niels \surname{L\"orch}}
\affiliation{Centre for Quantum Technologies, %
National University of Singapore, 3 Science Drive 2, %
Singapore 117543, Singapore}
\affiliation{Universit\"at Heidelberg, Philosophenweg 16-19, %
69120 Heidelberg, Germany}

\author{Jun \surname{Suzuki}}
\affiliation{National Institute of Informatics, 2-1-2 Hitotsubashi, %
Chiyoda-ku, Tokyo 101-8430, Japan}

\author{Berthold-Georg \surname{Englert}}
\affiliation{Centre for Quantum Technologies, %
National University of Singapore, 3 Science Drive 2, %
Singapore 117543, Singapore}
\affiliation{Department of Physics, National University of Singapore, %
2 Science Drive 3, Singapore 117542, Singapore}

\date{10 March 2011}

\begin{abstract}
A system of three spin-1/2 atoms allows the construction of a
reference-frame-free (RFF) qubit in the subspace with total angular momentum
$j=1/2$.  
The RFF qubit stays coherent perfectly as long as the spins of the three atoms
are affected homogeneously. 
The inhomogeneous evolution of the atoms causes decoherence, but this
decoherence can be suppressed efficiently by applying a bias magnetic field of
modest strength perpendicular to the plane of the atoms. 
The resulting lifetime of the RFF qubit can be many days, 
making RFF qubits of this kind promising candidates for quantum
information storage units. 
Specifically, we examine the situation of three $^6\textrm{Li}$ atoms trapped
in a $\textrm{CO}_2$-laser-generated optical lattice and find that, with
conservatively estimated parameters, a stored qubit maintains a fidelity of
0.9999 for two hours. 
\end{abstract}

\pacs{03.67.Pp, 03.65.Yz, 03.67.Lx}

\maketitle

\section{Introduction}
The processing of quantum information --- be it for quantum communication, for
quantum key distribution, or for quantum computation --- requires the
storage, manipulation, and retrieval of qubits that are carried by physical 
systems. 
This ``hardware'' can consist of selected well-controllable degrees of freedom
of single photons, trapped ions, N-V centers in solids, or quantum dots, to
name a few. 
All of them have advantages and disadvantages that make them well-fit for some
purposes and unsuitable for others~\cite{Ladd+5:10}. 
Practical applications that go beyond proof-of-principle experiments rely on
qubits that are sufficiently robust for the task at hand. 

For example, fault-tolerant quantum computation requires a gate fidelity that
is very close to unity. 
Lack of control over the system, however, always gives rise to decoherence. 
The typical decoherence time for quantum dots, ions in a trap, or diamond N-V
centers is in the order of microseconds or milliseconds, and a decoherence
time of a few seconds is in reach of the current technologies~\cite{Ladd+5:10}. 
This is still not quite sufficient for carrying out some complicated gate
operations, nor for storage purposes~\cite{Simon+23:10}. 

We explore here a scheme to overcome the decoherence problem with
reference-frame-free (RFF) qubits constructed from three spin-1/2 neutral
atoms. 
These RFF qubits have a remarkably long lifetime --- a NMR proof-of-principle
experiment that employs three spin-1/2 nuclei is on record~\cite{Viola+5:01} 
--- and the alignment of reference frames between observers, or the drift of
frame between storage and read-out, is not an issue. 

The said construction of RFF qubits from trios of spin-1/2 atoms was studied
previously~\cite{Bartlett+2:07,Suzuki+2:08}. 
These atoms are individually highly sensitive to magnetic stray fields, but
their symmetric RFF states are completely insensitive as long as the stray
field affects all three atoms in the same way~\cite{Lidar+1:03}. 
Decoherence of the RFF qubit could, therefore, result from spatial
inhomogeneities of the stray field. 
We investigate the effect of such inhomogeneities and demonstrate that the RFF
qubits are highly robust: For typical experimental parameters, the magnetic
stray fields are of no concern.

Rather, the lifetime of the RFF qubit is limited by the magnetic dipole-dipole
interaction among the constituent atoms in conjunction with intrinsic
imperfections of the experimental set-up.
Our analysis, which uses conservatively estimated parameters and reasonable
assumptions about experimental imperfections, shows that a stored qubit
maintains a fidelity of 0.9999 or 0.999 for two or seven hours, respectively.

The outline of the paper is as follows. 
We first review the construction of the RFF qubit in Sec.~\ref{sec:2}, 
followed by a qualitative discussion of  possible causes of decoherence. 
In Sec.~\ref{sec:3}, we analyze the robustness of the RFF qubit against
decoherence under random stray magnetic fields and conclude that the stray
fields are innocuous.
We then consider, in Sec.~\ref{sec:4}, the inhomogeneous magnetic dipole
fields of the partner atoms and find that, in view of unavoidable experimental
imperfections, they are the dominating effect that limits the period for which
quantum information can be stored.   
Section~\ref{sec:5} briefly argues that the RFF system formed by three
spin-1/2 constituents is more robust than the system formed by four spin-1/2
constituents. 
An experimental realization could use ultracold atoms in an optical potential
of a suitable geometry
\cite{Jaksch+4:98,Kohl+4:05,Cho:07,Trotzky+9:08}; 
we mention one possibility of creating a potential of this kind in
Sec.~\ref{sec:6}.  
We close with a Summary and Discussion.

\section{Construction of the RFF qubit}\label{sec:2}
For a system of three spin-1/2 particles, the total spin $j$ of the system can
be either 1/2 or 3/2. 
In the ${j=3/2}$ subspace, there are four states with different $m$ values; 
in the ${j=1/2}$ subspace, there are two different $m$ values, ${m=\pm1/2}$, 
with two states each, which we distinguish by the quantum number $\lambda$:
$\ket{j=1/2,m=\pm1/2,\lambda}$ with ${\lambda=0}$ or ${\lambda=1}$.

We construct these four orthogonal basis kets in the ${j=1/2}$ sector by a
variant of the procedure described in Ref.~\cite{Suzuki+2:08}.
When denoting the Pauli vector operator for the $k$th atom by
$\bvec{\sigma}_k$,
the total spin vector operator is given by
\begin{equation}\label{eq:3}
  \bvec{J}=\frac{1}{2}\sum_{k=1}^3\bvec{\sigma}_k
\end{equation}
in units of $\hbar$.
The lowering operator ${J_-=J_x-iJ_y}$ has the partner 
operators~\cite{err.Suzuki+2:08}
\begin{equation}
  \label{eq:3a}
  Q_0=\frac{1}{\sqrt{3}}\sum_{k=1}^3q^k\sigma_{k-}\,,\quad
  Q_1=\frac{1}{\sqrt{3}}\sum_{k=1}^3q^{-k}\sigma_{k-}\,,
\end{equation}
where ${\sigma_{k-}=(\sigma_{kx}-i\sigma_{ky})/2}$ is the lowering operator
for the $k$th atom, and ${q=e^{i2\pi/3}}$ is the basic cubic root of unity.
We write $\ket{\pm,\lambda}$ for $\ket{j=1/2,m=\pm1/2,\lambda}$ for brevity
and choose the ``$+$'' kets in accordance with
\begin{equation}
  \label{eq:def+}
  \ket{+,\lambda}=Q_{\lambda}\ket{\!\uparrow\uparrow\uparrow}\,,
\end{equation}
and an application of $J_-$ gives the corresponding ``$-$'' kets,
\begin{equation}
  \label{eq:def-}
  \ket{-,\lambda}=J_-\ket{+,\lambda}
                 =J_-Q_{\lambda}\ket{\!\uparrow\uparrow\uparrow}\,,
\end{equation}
with the outcomes
\begin{eqnarray}
  \label{eq:1}
  \ket{+,0}&=&\bigl(\ket{\!\downarrow\uparrow\uparrow}q
               +\ket{\!\uparrow\downarrow\uparrow}q^2
               +\ket{\!\uparrow\uparrow\downarrow}\bigr)/\sqrt{3}\,,
\nonumber\\  
  \ket{+,1}&=&\bigl(\ket{\!\downarrow\uparrow\uparrow}q^2
               +\ket{\!\uparrow\downarrow\uparrow}q
               +\ket{\!\uparrow\uparrow\downarrow}\bigr)/\sqrt{3}
\end{eqnarray}
and
\begin{eqnarray}
  \label{eq:2}
  \ket{-,0}&=&-\bigl(\ket{\!\uparrow\downarrow\downarrow}q
                +\ket{\!\downarrow\uparrow\downarrow}q^2
                +\ket{\!\downarrow\downarrow\uparrow}\bigr)/\sqrt{3}\,,
\nonumber\\
  \ket{-,1}&=&-\bigl(\ket{\!\uparrow\downarrow\downarrow}q^2
                +\ket{\!\downarrow\uparrow\downarrow}q
                +\ket{\!\downarrow\downarrow\uparrow}\bigr)/\sqrt{3}\,.
\end{eqnarray}
The arrows symbolize ``spin up'' and ``spin down'' in the $z$-direction,
so that ${\ket{\!\uparrow\uparrow\uparrow}=\ket{j=3/2,m=1/2}}$ in
Eqs.~(\ref{eq:def+}) and (\ref{eq:def-}).
The $\ket{\pm,\lambda}$ kets have the usual
properties as eigenstates of $J_z$, namely
\begin{eqnarray}
  \label{eq:J-applied}
  J_x\ket{\pm,\lambda}&=&\ket{\mp,\lambda}/2\,,\nonumber\\
  J_y\ket{\pm,\lambda}&=&\ket{\mp,\lambda}(\pm i/2)\,,\nonumber\\
  J_z\ket{\pm,\lambda}&=&\ket{\pm,\lambda}(\pm1/2)\,,
\end{eqnarray}
as one verifies immediately.

Since the eigenvalues of $\bvec{J}^2$ distinguish the ${j=1/2}$ and ${j=3/2}$
sectors, we can express the respective projectors in terms of $\bvec{J}^2$,
\begin{equation}\label{eq:projectors}
  P_{j=1/2}=\frac{5}{4}-\frac{1}{3}{\bvec{J}}^2\,,\quad
  P_{j=3/2}=\frac{1}{3}{\bvec{J}}^2-\frac{1}{4}\,,
\end{equation}
consistent with ${1=P_{j=1/2}+P_{j=3/2}}$ and
$\bvec{J}^2=\frac{3}{4}P_{j=1/2}+\frac{15}{4}P_{j=3/2}$. 
Alternatively, the projector onto the subspace with ${j=1/2}$ is given by
\begin{eqnarray}
P_{j=1/2}&=&\sum_{\sigma=\pm}\sum_{\lambda=0,1}
            \ket{\sigma,\lambda}\bra{\sigma,\lambda}\nonumber\\
&=&\frac{1}{6}(3-\bvec{\sigma}_1\cdot\bvec{\sigma}_2
                -\bvec{\sigma}_2\cdot\bvec{\sigma}_3
                -\bvec{\sigma}_3\cdot\bvec{\sigma}_1)\,,\label{P1/2}
\end{eqnarray}
where the latter expression is available either as a consequence of
Eqs.~(\ref{eq:1}) and (\ref{eq:2}) or of Eqs.~(\ref{eq:projectors}) and
(\ref{eq:3}). 

The ${j=1/2}$ subspace and the ${j=3/2}$ subspace are both four-dimensional
Hilbert spaces and, therefore, they can be regarded as tensor product spaces
of two qubits, respectively. 
This is of no consequence for the ${j=3/2}$ sector, but it permits to write the
${j=1/2}$ sector as composed of a rotationally invariant signal qubit --- the
RFF qubit --- and an idler qubit~\cite{Suzuki+2:08}:
the kets $\ket{\sigma,\lambda}$ of Eqs.~(\ref{eq:def+})--(\ref{eq:2})
are labeled by the idler quantum number ${\sigma=\pm}$ and the signal quantum
number ${\lambda=0,1}$.

With the idler qubit in a maximally mixed state, the RFF state is identified
by
\begin{equation}
\rho^{\mathrm{RFF}}=\sum_{\sigma=\pm}\sum_{\lambda,\lambda'=0,1}
                   \ket{\sigma,\lambda}\frac{1}{2}\rho_{\lambda\lambda'}
                   \bra{\sigma,\lambda'}
=\frac{1}{2}1_2\otimes\tilde\rho^{\mathrm{RFF}}\,, \label{rhorff}
\end{equation}
where this tensor-product structure applies within the subspace with ${j=1/2}$.
We exhibit the statistical operator of the signal qubit alone,
\begin{equation}\label{eq:encoding}
\tilde\rho^{\mathrm{RFF}}=\sum_{\lambda,\lambda'=0,1}
                        \ket{\lambda}\rho_{\lambda\lambda'}\bra{\lambda'}\,,
\end{equation}
by tracing over the idler qubit.
The hermitian Pauli operators $\Sigma_1$, $\Sigma_2$, $\Sigma_3$ for the RFF
qubit,
\begin{eqnarray}
  \label{eq:Sigma}
  \frac{1}{2}\bigl(\Sigma_1+i\Sigma_2\bigr)&=&
  \sum_{\sigma=\pm}\ket{\sigma,0}\bra{\sigma,1}=
   1_2\otimes\bigl(\ket{0}\bra{1}\bigr)\,,
  \nonumber\\
\Sigma_3&=&
  \sum_{\sigma=\pm}\sum_{\lambda=0,1}\ket{\sigma,\lambda}(-1)^\lambda
                                  \bra{\sigma,\lambda}\nonumber\\&=&
  1_2\otimes{\bigl(\ket{0}\bra{0}-\ket{1}\bra{1}\bigr)}\,,
\end{eqnarray}
are explicitly given by
\begin{eqnarray}
\Sigma_1+i\Sigma_2&=&\frac{1}{3}(\bvec{\sigma}_1\cdot\bvec{\sigma}_2
                       +q^2\bvec{\sigma}_2\cdot\bvec{\sigma}_3
                       +q\bvec{\sigma}_3\cdot\bvec{\sigma}_1)\,,
\nonumber\\
\Sigma_3&=&\frac{1}{\sqrt{12}}\bvec{\sigma}_1
            \cdot(\bvec{\sigma}_2\times\bvec{\sigma}_3)\,.\label{Sigma}
\end{eqnarray}
These are clearly rotationally invariant and possess the algebraic properties
of Pauli spin operators in the ${j=1/2}$ subspace, 
such as ${(\Sigma_1)^2=P_{j=1/2}}$ and ${\Sigma_1\Sigma_2=i\Sigma_3}$. 

Once the information is encoded in a RFF qubit (\ref{eq:encoding}), 
with the idler in the maximally mixed state as in Eq.~(\ref{rhorff}) or in some
other state, 
the information will be perfectly preserved as long as all three spin-1/2 atoms
precess in unison. 
In the non-ideal circumstances of a real experimental situation, however,
the interaction with the environment and the interactions among the physical
carriers of the qubit could cause decoherence, because the atoms may be
subject to torques of different strengths.
Inevitably, there will be sources of noise over which the experimenter lacks
control. 
It is our first objective to demonstrate that the information stored
in the RFF qubit is preserved for a long time if magnetic stray fields with
typical properties affect the carrier atoms.

\section{Magnetic stray fields}\label{sec:3}
\subsection{Noise model}\label{sec:3A}
The part of the Hamiltonian that describes the effect of the noisy magnetic
field on the trio of atoms is given by 
\begin{equation}\label{eq:3-0}
H_{\mathrm{noise}}(t)
=\muB\sum_{k=1}^3{\bvec{b}}_k(t)\cdot\bvec{\sigma}_k\,,
\end{equation}
where $\mu_{\textsc{b}}$ is the Bohr magneton (if necessary multiplied by a
gyromagnetic ratio) and $\bvec{b}_k(t)$ is the randomly fluctuating 
magnetic stray field that acts on the $k$th atom.
The  $\bvec{b}_k(t)$s vanish on average,
\begin{equation}
  \label{eq:3-1}
  \overline{\,\bvec{b}_k(t)\,}=0\,,
\end{equation}
where the overline indicates the stochastic average.
Since the atoms are close to each other, the fluctuations in the magnetic
fields at the positions of the atoms are not independent but correlated. 
The dominant part of the noisy magnetic field is the same for all three atoms,
and only a small part of the noise affects the atoms differently as a
consequence of the nonzero gradient of the magnetic field. 
Upon denoting the gradient dyadic of the magnetic stray field by
$\mathbf{G}(t)$,
we have
\begin{equation}
  \label{eq:3-2}
  \bvec{b}_k(t)-\bvec{b}_l(t)
  =\mathbf{G}(t)\cdot(\bvec{r}_k-\bvec{r}_l)\,,
\end{equation}
where $\bvec{r}_k$ is the position vector for the $k$th atom, and
$\mathbf{G}(t)$ is assumed to be independent of position within the small
volume of relevance.
This gradient component is the inhomogeneous noise that gives rise to
decoherence of the RFF qubit, while the homogeneous noise is innocuous. 

In the noise model considered, every component of the homogeneous stray field
and every component of the gradient dyadic has a random
gaussian distribution with a vanishing mean. 
Owing to the Maxwell's equations, the gradient dyadic has to be symmetric and
traceless: 
\begin{eqnarray}
&\bvec{v}_1 \cdot \mathbf{G}(t) \cdot\bvec{v}_2
  =\bvec{v} _2 \cdot \mathbf{G}(t) \cdot\bvec{v}_1\,,&\nonumber\\
&\displaystyle
\sum_{a=x,y,z}\bvec{e}_a \cdot \mathbf{G}(t) \cdot\bvec{e}_a=0
\,.&
\end{eqnarray}
It follows that the two-time correlation function of the field gradient has the
form 
\begin{eqnarray}
&&\overline{\,\bvec{v}_1 \cdot \mathbf{G}(t) \cdot \bvec{v}_2 
    \;\bvec{v}_3 \cdot \mathbf{G}(t') \cdot \bvec{v}_4\,}\nonumber\\
&=&(3\bvec{v}_1\cdot\bvec{v}_3 \;\bvec{v}_2\cdot\bvec{v}_4
   +3\bvec{v}_1\cdot\bvec{v}_4\; \bvec{v}_2\cdot\bvec{v}_3
   -2\bvec{v}_1\cdot\bvec{v}_2\; \bvec{v}_3\cdot\bvec{v}_4)
\nonumber\\
&&\times \frac{1}{4}g^2 e^{-\Gamma|t'-t|}\label{G}
\end{eqnarray}
for any four vectors $\bvec{v}_1$, $\bvec{v}_2$, $\bvec{v}_3$,
and $\bvec{v}_4$ that pick out the components of $\mathbf{G}$, whereby
$g^2$ is the variance of the gaussian distribution of the
diagonal entries of $\mathbf{G}$, and $\Gamma$ is the decay constant for the
temporal correlation.
We note that the diagonal elements and the off-diagonal elements of the
gradient matrix do not have the same variance: 
\begin{eqnarray}\label{Gon-Goff}
\overline{\,\bvec{e} \cdot \mathbf{G}(t) \cdot \bvec{e} 
    \;\bvec{e} \cdot \mathbf{G}(t') \cdot \bvec{e}\,}
&=&\frac{4}{3}
\overline{\,\bvec{e} \cdot \mathbf{G}(t) \cdot \bvec{e}' 
    \;\bvec{e} \cdot \mathbf{G}(t') \cdot \bvec{e}'\,}
\nonumber\\&=&g^2e^{-\Gamma|t'-t|}\,,
\end{eqnarray}
where $\bvec{e}$ and $\bvec{e}'$ are two orthogonal unit vectors.
The autocorrelation function for the diagonal components of $\mathbf{G}$ is 4/3
times that of the off-diagonal components, while they have the same correlation
time $1/\Gamma$.

As a consequence of Eq.~(\ref{G}), the autocorrelation of the field difference
of Eq.~(\ref{eq:3-2}) is
\begin{eqnarray}
  \label{eq:3-3}
&&\overline{\,
  \bvec{v}_1\cdot\bigl(\bvec{b}_k(t)-\bvec{b}_l(t)\bigr)\;
\bigl(\bvec{b}_k(t')-\bvec{b}_l(t')\bigr)\cdot\bvec{v}_2\,}
\nonumber\\&=&\frac{1}{4}g^2 e^{-\Gamma|t'-t|}
[3\bvec{v}_1\cdot\bvec{v}_2\;{{\bvec{r}}_{kl}}^2
 +\bvec{v}_1\cdot{\bvec{r}}_{kl}\;
  {\bvec{r}}_{kl}\cdot\bvec{v}_2]\qquad
 \end{eqnarray}
with ${\bvec{r}_{kl}=\bvec{r}_k-\bvec{r}_l}$.
The correlation between the stray fields at the sites of the $k$th atom and
the $l$th atom is then given by
\begin{eqnarray}
  \label{eq:3-4}
&&\overline{\,
  \bvec{v}_1\cdot\bvec{b}_k(t)\;
  \bvec{b}_l(t')\cdot\bvec{v}_2\,}
=e^{-\Gamma|t'-t|}\Bigl[b^2\bvec{v}_1\cdot\bvec{v}_2\nonumber\\
 &&\hspace*{3em}\mbox{}
-\frac{1}{8}g^2(3\bvec{v}_1\cdot\bvec{v}_2\;
  {\bvec{r}_{kl}}^2+\bvec{v}_1\cdot\bvec{r}_{kl}\;
  \bvec{r}_{kl}\cdot\bvec{v}_2)\Bigr]\,,\qquad
\end{eqnarray}
where ${b^2\gg g^2 {\bvec{r}_{kl}}^2}$ is the strength of the
same-site correlation, as we recognize by a look at the ${k=l}$ version,
\begin{equation}
  \label{eq:3-5}
  \overline{\,
  \bvec{v}_1\cdot\bvec{b}_k(t)\;
  \bvec{b}_k(t')\cdot\bvec{v}_2\,}
          =b^2e^{-\Gamma|t'-t|}\bvec{v}_1\cdot\bvec{v}_2\,.
\end{equation}
Consistency with Eq.~(\ref{eq:3-3}) is established by using Eq.~(\ref{eq:3-4})
four times on the right-hand side of 
\begin{eqnarray}
  \label{eq:3-6}
  &&\overline{\,
  \bvec{v}_1\cdot\bigl(\bvec{b}_k(t)-\bvec{b}_l(t)\bigr)\;
\bigl(\bvec{b}_k(t')-\bvec{b}_l(t')\bigr)\cdot\bvec{v}_2\,}
\nonumber\\&=&
  \overline{\,\bvec{v}_1\cdot\bvec{b}_k(t)\;
  \bvec{b}_k(t')\cdot\bvec{v}_2\,}
+  \overline{\,\bvec{v}_1\cdot\bvec{b}_l(t)\;
  \bvec{b}_l(t')\cdot\bvec{v}_2\,}\nonumber\\&&\mbox{}
-  \overline{\,\bvec{v}_1\cdot\bvec{b}_k(t)\;
  \bvec{b}_l(t')\cdot\bvec{v}_2\,}
-  \overline{\,\bvec{v}_1\cdot\bvec{b}_l(t)\;
  \bvec{b}_k(t')\cdot\bvec{v}_2\,}\,.\qquad
\end{eqnarray}

Equations (\ref{eq:3-1}) and (\ref{eq:3-4}) define the noise model that we use
in Sec.~\ref{sec:3C} below to derive the master equation by which we then
study the effect of the inhomogeneous magnetic stray fields on the RFF qubit
in Sec.~\ref{sec:3D}, and on a single-atom qubit in Sec.~\ref{sec:3E}.  
The model is characterized by the three parameters $\Gamma$, $b^2$, and
$g^2$, which would have to be determined from experimental data when
applying the model to an actual laboratory situation.
Other noise models are conceivable, in particular if one wants to describe a
specific noise source of known characteristics. 
The noise model of Eqs.~(\ref{eq:3-1}) and (\ref{eq:3-4}) is generic, however,
and quite suitable for the purpose at hand.  

In an experimental realization of the three-atom RFF qubit, there will be
nearby Helmholtz coils for the precise control of the magnetic field at the
location of the atoms.
Typically, these coils are about $50\,\mathrm{cm}$ away and carry currents of
about $1\,\mathrm{A}$ that are stabilized to $100\,\mathrm{ppm}$ or
better~\cite{Gleb-PC:10}.
Now, a current of $0.1\,\mathrm{mA}$ at a distance of $0.5\,\mathrm{m}$ gives
rise to a magnetic field of $4\times10^{-11}\,\mathrm{T}$ and a field gradient
of $8\times10^{-11}\,\mathrm{T}/\mathrm{m}$.
Not assuming any fortunate cancellation of the contributions from different
coils, the values ${b=5\times10^{-10}\,\mathrm{T}}$ and 
${g=10^{-9}\,\mathrm{T}/\mathrm{m}}$ are conservative estimates for these
noise parameters.
The temporal properties of the fluctuating currents tend to be dominated by
the ubiquitous $50\,\mathrm{Hz}$ noise that the wires pick up, while
high-frequency noise can be filtered out very efficiently, so that a
correlation time of ${1/\Gamma=20\,\mathrm{ms}}$ is a reasonable
estimate~\cite{Gleb-PC:10}.
We will use these numbers throughout the paper.

\subsection{Lithium-6}\label{sec:3B}
\begin{figure}
\centerline{\includegraphics{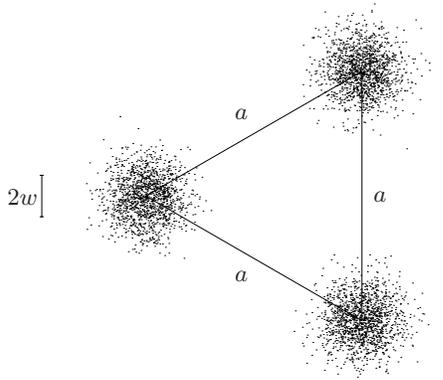}}
\caption{\label{fig:triangle}%
Three $^6$Li atoms are trapped at the corners of an equilateral triangle. 
The probability clouds indicate the center-of-mass distributions whose spread
$w$ is about one-twelfth of $a$, the distance between the atoms.
}
\end{figure}
\begin{figure}
\centerline{\includegraphics{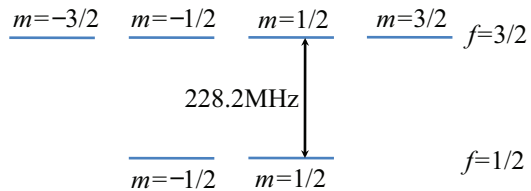}}  
\caption{\label{fig:Li}%
Ground-state hyperfine levels of the neutral $^6$Li atom.
The ${f=3/2}$ quartet is separated from the ${f=1/2}$ doublet by a transition
frequency of ${228.2\,\mathrm{MHz}}$.  
Three $^6$Li atoms confined to their ${f=1/2}$ ground states serve as the
spin-1/2 particles from which the RFF qubit is constructed. 
}
\end{figure}

To be specific, but also mindful of possible experiments with two-dimensional
confinement \cite{Lee+4:09}, we consider the situation of
Fig.~\ref{fig:triangle}: Three $^6$Li atoms
at the corners of an equilateral triangle, perhaps the sites of neighboring
minima of an optical potential such as the one discussed below in
Sec.~\ref{sec:5}. 
Each $^6$Li atom is in the
hyperfine ground state with ${f=1/2}$, which is energetically below the
${f=3/2}$ hyperfine state by ${2\pi\hbar\times228.2\,\mathrm{MHz}}$; see
Fig.~\ref{fig:Li}. 

We denote the electronic spin operator of the $k$th atom by 
${\bvec{s}}_k$, with ${{{\bvec{s}}_k}^2=3\hbar^2/4}$, so that the
energy of the atom trio in an external homogeneous magnetic bias field
${\bvec{b}}_0$ is given by 
\begin{equation}
  \label{eq:a}
  H_{\mathrm{bias}}=\sum_{k=1}^3
                   2\frac{\muB}{\hbar}{\bvec{s}}_k
                   \cdot{\bvec{b}}_0\,,
\end{equation}
where we take the value of $2$ for the gyromagnetic factor of the electron.
With each atom confined to its ${f=1/2}$ ground state, this becomes
\begin{equation}
  \label{eq:b}
    H_{\mathrm{bias}}=\sum_{k=1}^3
                   {\left(-\frac{2}{3}\right)}
                   \frac{\mu_{\mathrm{B}}}{\hbar}{\bvec{F}}_k
                   \cdot{\bvec{b}}_0\,,
\end{equation}
where ${-2/3}$ is the gyromagnetic ratio and 
${{\bvec{F}}_k=(\hbar/2)\bvec{\sigma}_k}$ is the total atomic angular
momentum in the present context; the magnetic moment of the spin-1 nucleus is
ignored. 
As indicated, here we identify the Pauli operators of Eqs.~(\ref{eq:3}),
(\ref{P1/2}), (\ref{Sigma}), or (\ref{eq:3-0}). 

We choose the bias field in the $z$-direction, perpendicular to the
$xy$-plane in which the atoms are located, 
${{\bvec{b}}_0=-B_0{\bvec{e}}_z}$, and express its strength in terms
of the circular frequency $\omega_0$: ${\hbar\omega_0=2\muB B_0/3}$;
then
\begin{equation}
  \label{eq:c}
  H_{\mathrm{bias}}=\frac{1}{2}\hbar\omega_0\sum_{k=1}^3\sigma_{kz}
                 =\hbar\omega_0 J_z\,.
\end{equation}
The coupling of the ${f=1/2}$ and the ${f=3/2}$ multiplets by the bias field
is ignored, which is permissible if the field is weak on the scale set by the
energy difference, that is: ${\omega_0\ll2\pi\times228.2\,\mathrm{MHz}}$.
For example, this condition is met for the modest field strength of
${B_0=2\,\mathrm{mG}=2\times10^{-7}\,\mathrm{T}}$, when
$\omega_0=2\pi\times{2\,\mathrm{kHz}}$ 
is a thousandth of a percent of the transition
frequency, and transition probabilities are of the order of
${(10^{-5})^2=10^{-10}}$. 

We need the bias field to fight the ``internal magnetic pollution'' that
originates in the magnetic dipole-dipole interaction between the spin-1/2
atoms.
In terms of the electronic spin operators, this interaction energy 
is~\cite{ContactTerm}
\begin{equation}
  \label{eq:d}
  H_{\mathrm{dd}}=\frac{\mu_0}{4\pi}
                {\left(\frac{2\muB}{\hbar}\right)}^2
                \frac{1}{a^3}
                \sum_{(k,l)}\left({\bvec{s}}_k\cdot{\bvec{s}}_l
                             -3{\bvec{s}}_k\cdot{\bvec{e}}_{kl}\,
                               {\bvec{e}}_{kl}\cdot{\bvec{s}}_l
                            \right)\,,
\end{equation}
where ${a=|{\bvec{r}}_{kl}|}$ is the common distance between the atoms 
at the corners of the equilateral triangle, 
${\bvec{e}}_{kl}={\bvec{r}}_{kl}/a$ is the unit vector that points
from the $k$th to the $j$th atom, and the summation is over the three pairs.
As in the transition from Eq.~(\ref{eq:a}) to Eq.~(\ref{eq:b}), the restriction
to the ${f=1/2}$ ground state amounts to the replacement
\begin{equation}
  \label{eq:e}
{\bvec{s}}_k\to-\frac{1}{3}{\bvec{F}}_k
=-\frac{\hbar}{6}\bvec{\sigma}_k\,,  
\end{equation}
which turns Eq.~(\ref{eq:d}) into
\begin{equation}
  \label{eq:f}
  H_{\mathrm{dd}}=\frac{1}{3}\hbar\Omega\sum_{(k,l)}
                 \left({\bvec{\sigma}}_k\cdot{\bvec{\sigma}}_l
                       -3{\bvec{\sigma}}_k\cdot{\bvec{e}}_{kl}\,
                        {\bvec{e}}_{kl}\cdot{\bvec{\sigma}}_l
                   \right)
\end{equation}
with
\begin{equation}
  \label{eq:g}
  \hbar\Omega=\frac{\mu_0}{4\pi}
                \frac{{\muB}^2}{3a^3}\,.
\end{equation}
For a distance of ${a=883\,\mathrm{nm}}$ (see Sec.~\ref{sec:6} below), 
we have ${\Omega=2\pi\times6\,\mathrm{mHz}}$, smaller than $\omega_0$ by a
factor of $3\times10^{5}$, 
so that the transitions induced by $H_{\mathrm{dd}}$ are
completely suppressed in the presence of a $2\,\mathrm{mG}$ bias field.
Therefore, only the part of $H_{\mathrm{dd}}$ that commutes with
$H_{\mathrm{bias}}$ of Eq.~(\ref{eq:c}) is relevant, and we arrive at
\begin{equation}
  \label{eq:h}
  H_{\mathrm{dd}}=\frac{1}{3}\hbar\Omega\left(3J_z^2-{\bvec{J}}^2\right)
\end{equation}
as the effective Hamiltonian for the magnetic dipole-dipole interaction among
the atoms.
Note that this $H_{\mathrm{dd}}$ vanishes in the ${j=1/2}$ sector where the
signal and idler qubits reside.

\begin{figure}
\centerline{\includegraphics{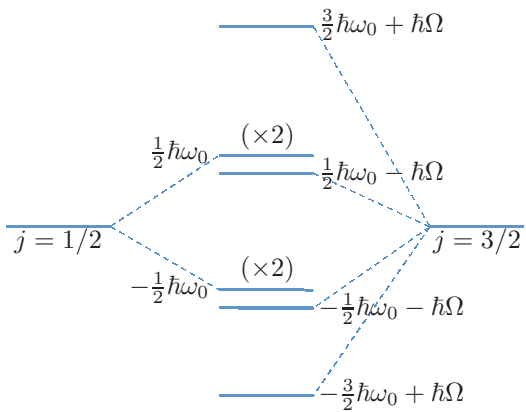}}
\caption{\label{fig:threeAtoms}%
Level scheme for the effective three-atom Hamiltonian of Eq.~(\ref{eq:i}). 
The separations are not drawn to scale: $\hbar\omega_0$ is many orders of
magnitude larger than $\hbar\Omega$.
The two ${j=1/2}$ levels are degenerate doublets; this degeneracy is exploited
for the encoding of the robust signal qubit. 
}
\end{figure}

The combined effective Hamiltonian 
\begin{equation}
  \label{eq:i}
  H_{\mathrm{bias}}+H_{\mathrm{dd}}= \hbar\omega_0J_z^{\ }
           +\frac{1}{3}\hbar\Omega{\left(3J_z^2-{\bvec{J}}^2\right)}
\end{equation}
has the non-degenerate  eigenvalues
${\pm\frac{3}{2}\hbar\omega_0+\hbar\Omega}$ and  
${\pm\frac{1}{2}\hbar\omega_0-\hbar\Omega}$ in the ${j=3/2}$ sector,
and the two-fold eigenvalues $\pm\frac{1}{2}\hbar\omega_0$ 
in the ${j=1/2}$ sector; see Fig.~\ref{fig:threeAtoms}.
The energy differences correspond to
transition frequencies of about ${1,2,3}\times\omega_0/(2\pi)$, which are in
the few-kHz range, and to transition frequency $\Omega/(2\pi)$,
which is $6\,\mathrm{mHz}$, if we continue to use the numbers found above.
There is a clear separation of time scales, then, and
noise with a correlation time $1/\Gamma$ in the order of $20\,\mathrm{ms}$ ---
which we regard as a typical number, see above --- would
not be able to induce transitions between the states separated by several
$\hbar\omega_0$ while it will mix the states that are separated by
$\hbar\Omega$ only or not at all.   

There are, of course, stray fields in the radio frequency range but their
sources (the radio stations) are far away so that the gradient
parameter $g$ is extremely small, and noise of this kind is of no concern.
By contrast, noise originating in nearby sources --- 
current carrying wires in the vicinity of the laboratory, say --- is relevant.

\subsection{Master equation}\label{sec:3C}
In view of this separation of time scales --- very fast
$\omega_0$-oscillations and very slow $\Omega$-oscillations on the scale
set by the correlation time $1/\Gamma$ of the random stray field --- we can
use master equation techniques to account for the net effect of the noise.
For the purpose of deriving the Lind\-blad operators of the master equation, we
put $H_{\mathrm{dd}}$ of Eq.~(\ref{eq:i}) aside and use an interaction picture
in which the fast $\omega_0$-oscillations of $H_{\mathrm{bias}}$ are
transformed away.
The statistical operator in this interaction picture, denoted by
$\IP{\varrho}(t)$, then obeys the von Neumann equation of motion
\begin{equation}
  \label{eq:ME1}
  \frac{\partial}{\partial t}\IP{\varrho}(t)
=\frac{i}{\hbar}\bigl[\IP{\varrho}(t),\IP{H}_{\mathrm{noise}}(t)\bigr]
\end{equation}
with
\begin{eqnarray}
  \label{eq:ME2}
  \IP{H}_{\mathrm{noise}}(t)&=&
e^{i H_{\mathrm{bias}}t/\hbar}{H}_{\mathrm{noise}}(t)
e^{-i H_{\mathrm{bias}}t/\hbar}
\nonumber\\&=&
e^{i\omega_0tJ_z}\Bigl(-\frac{\muB}{3}\Bigr)
\sum_{k=1}^3\bvec{b}_k(t)\cdot\bvec{\sigma}_ke^{-i\omega_0tJ_z},\qquad
\end{eqnarray}
where the replacement ${\muB\to-\muB/3}$ accounts for the gyromagnetic
ratio that we first met in the transition from Eq.~(\ref{eq:a}) to
Eq.~(\ref{eq:b}). 

The unitary evolution operator $U(T)$ links $\IP{\varrho}(T)$ to the initial
statistical operator $\IP{\varrho}(0)$,
\begin{equation}
  \label{eq:ME3}
  \IP{\varrho}(T)=U(T)\IP{\varrho}(0)U(T)^{\dagger}\,.
\end{equation}
We solve the Lippmann--Schwinger equation
\begin{equation}
  \label{eq:ME4}
  U(T)=1-\frac{i}{\hbar}\int_0^T\!\!dt\, \IP{H}_{\mathrm{noise}}(t)U(t)
\end{equation}
to second order in ${H}_{\mathrm{noise}}$,
\begin{equation}
  \label{eq:ME5}
  U(T)\simeq1-i\phi_1(T)-\frac{1}{2}\phi_1(T)^2-i\phi_2(T)\,,
\end{equation}
where the hermitian phases $\phi_1(T)$ and $\phi_2(T)$ are given by
\begin{eqnarray}
  \label{eq:ME7}
  \phi_1(T)&=&\frac{1}{\hbar}\int_0^T\!\!dt\,\IP{H}_{\mathrm{noise}}(t)\,,
\nonumber\\
  \phi_2(T)&=&\frac{1}{2i\hbar^2}\int_0^T\!\!dt\int_0^t\!\!dt'\,
              \bigl[\IP{H}_{\mathrm{noise}}(t),\IP{H}_{\mathrm{noise}}(t')\bigr]\,.
\qquad
\end{eqnarray}

To second order in ${H}_{\mathrm{noise}}$, then, we have
\begin{eqnarray}
  \label{eq:ME8}
  \IP{\varrho}(T)&=&\IP{\varrho}(0)+i[\IP{\varrho}(0),\phi_1(T)+\phi_2(T)]
\nonumber\\&&\mbox{}
+\frac{1}{2}\bigl[\phi_1(T),[\IP{\varrho}(0),\phi_1(T)]\bigr]\,,
\end{eqnarray}
and the stochastic averaging of Sec.~\ref{sec:3A} turns this into
\begin{eqnarray}
  \label{eq:ME9}
  \IP{\rho}(T)&=&\IP{\rho}(0)+i[\IP{\rho}(0),\overline{\,\phi_2(T)\,}]
\nonumber\\&&\mbox{}
+\frac{1}{2}\overline{\,\bigl[\phi_1(T),[\IP{\rho}(0),\phi_1(T)]\bigr]\,},
\end{eqnarray}
where ${\rho(t)=\overline{\,\varrho(t)\,}}$ and the initial statistical
operator is not affected by the averaging or the transition to the interaction
picture: 
${\IP{\varrho}(0)=\overline{\,\IP{\varrho}(0)\,}=\IP{\rho}(0)=\rho(0)}$.
Note that ${\overline{\,\phi_1(T)\,}=0}$ follows from Eq.~(\ref{eq:3-1}).

\begin{widetext} 
We take a closer look at the ``sandwich term'' in the double commutator,
\begin{equation}
  \label{eq:ME10}
\overline{\,\phi_1(T)\IP{\rho}(0)\phi_1(T)\,}
=\Bigl(\frac{\muB}{3\hbar}\Bigr)^2
\sum_{k,l=1}^3\int_0^T\!\!dt\int_0^T\!\!dt'\,
\IP{\bvec{\sigma}}_k(t)\cdot
\overline{\,\bvec{b}_k(t)\IP{\rho}(0)\bvec{b}_l(t')\,}
\cdot\IP{\bvec{\sigma}}_l(t')\,.
\end{equation}
Here, $T$ is much longer than the correlation time of the noise, 
${\Gamma T\gg1}$, so that there are very many cycles of the
$\omega_0$-oscillation in a short $t$-interval.
Therefore, the rapidly oscillating terms in $\IP{\bvec{\sigma}}_k(t)$ do not
contribute to the $t$-integration and the replacement
\begin{equation}
  \label{eq:ME11}
  \IP{\bvec{\sigma}}_k(t)=e^{i\omega_0tJ_z}\bvec{\sigma}_ke^{-i\omega_0tJ_z}
\to\bvec{\sigma}_k\cdot\bvec{e}_z\,\bvec{e}_z=\sigma_{kz}\bvec{e}_z
\end{equation}
is permissible; and likewise for $\IP{\bvec{\sigma}}_l(t')$.
This ``rotating-wave approximation'' takes us to
\begin{eqnarray}
  \label{eq:ME12}
\overline{\,\phi_1(T)\IP{\rho}(0)\phi_1(T)\,}
&=&\Bigl(\frac{\muB}{3\hbar}\Bigr)^2\sum_{k,l=1}^3
\sigma_{kz}\IP{\rho}(0)\sigma_{lz}
\int_0^T\!\!dt\int_0^T\!\!dt'\,
\overline{\,\bvec{e}_z\cdot\bvec{b}_k(t)\,\bvec{b}_l(t')\cdot\bvec{e}_z\,}
\nonumber\\
&=&\Bigl(\frac{\muB}{3\hbar}\Bigr)^2\sum_{k,l=1}^3
\sigma_{kz}\IP{\rho}(0)\sigma_{lz}
\Bigl[b^2-\frac{3}{8}(ga)^2(1-\delta_{kl})\Bigr]
\int_0^T\!\!dt\int_0^T\!\!dt'\,e^{-\Gamma|t-t'|}
\end{eqnarray}
\end{widetext}
after using Eq.~(\ref{eq:3-4}) for
${\bvec{v}_1=\bvec{v}_2=\bvec{e}_z}$,
${\bvec{r}_{kl}\cdot\bvec{e}_z=0}$, and ${{r_{kl}}^2=(1-\delta_{kl})a^2}$.
For ${\Gamma T\gg1}$, the remaining double integral equals $2T/\Gamma$, and we
arrive at
\begin{equation}
  \label{eq:ME13}
  \overline{\,\phi_1(T)\IP{\rho}(0)\phi_1(T)\,}
=\frac{T}{4\tau}\sum_{k=1}^3\sigma_{kz}\IP{\rho}(0)\sigma_{kz}
 +\frac{T}{\tau'}J_z\IP{\rho}(0)J_z
\end{equation}
with the time constants
\begin{equation}
  \label{eq:ME14a}
  \tau=3\Gamma\Bigl(\frac{\hbar}{\muB ga}\Bigr)^2
\end{equation}
and
\begin{equation}
  \label{eq:ME14b}
  \tau'=\frac{3(ga)^2}{8b^2-3(ga)^2}\tau\,.
\end{equation}
Since ${ga\ll b}$, we have ${\tau'\ll\tau}$, and the numbers of
Sec.~\ref{sec:3B}, that is: ${1/\Gamma=20\,\mathrm{ms}}$ and
${ga=9\times10^{-16}\,\mathrm{T}}$, 
give ${\tau=2\times10^{10}\,\mathrm{s}}$ --- an amazingly long time.
 
The replacement of Eq.~(\ref{eq:ME11}) gives a vanishing commutator in the
double integral for $\phi_2(T)$ in Eq.~(\ref{eq:ME7}), so that
${\overline{\,\phi_2(T)\,}=0}$ in Eq.~(\ref{eq:ME9}). 
In summary, then, we have
\begin{equation}
  \label{eq:ME15}
  \IP{\rho}(T)\simeq\IP{\rho}(0)+T\mathcal{L}\IP{\rho}(0)
\end{equation}
with the Lind\-blad operator $\mathcal{L}$ given by
\begin{equation}
  \label{eq:ME16}
  \mathcal{L}\rho=
\frac{1}{8\tau}\sum_{k=1}^3\bigl[\sigma_{kz},[\rho,\sigma_{kz}]\bigr]
 +\frac{1}{2\tau'}\bigl[J_z,[\rho,J_z]\bigr]\,,
\end{equation}
and the master equation in the interaction picture is simply
\begin{equation}
  \label{eq:ME17}
  \frac{\partial}{\partial t}\IP{\rho}(t)=\mathcal{L}\IP{\rho}(t)\,.
\end{equation}
Upon getting out of the interaction picture, and re-introducing the slow
$\Omega$-oscillations of $H_{\mathrm{dd}}$, this gives us the master equation
\begin{equation}
  \label{eq:ME18}
    \frac{\partial}{\partial t}\rho(t)=
  \frac{i}{\hbar}[\rho(t),H_{\mathrm{bias}}+H_{\mathrm{dd}}]+\mathcal{L}\rho(t)
\end{equation}
for the evolution of the coarse-grain, stochastically averaged, statistical
operator $\rho(t)$.

We note in passing that this master equation could alternatively be derived
with standard textbook methods, such as those that proceed from the Redfield
equation, here:
\begin{equation}\label{eq:ME18'}
  \frac{\partial}{\partial T}\IP{\rho}(T)=
\frac{1}{\hbar^2}\int_0^T\!\!dt\,
\overline{\,\bigl[\IP{H}_{\mathrm{noise}}(T),
            [\IP{\varrho}(t),\IP{H}_{\mathrm{noise}}(t)]\bigr]\,}\,,
\end{equation}
and invoke the Born--Markov approximation and the rotating-wave approximation
to arrive at Eq.~(\ref{eq:ME17}).
For details of this procedure, see section 3.2 
in Ref.~\cite{Breuer+Petruccione:02}, for example.

In view of the diagonal form of the Lind\-blad operator in
Eq.~(\ref{eq:ME16}), the expectation value ${\expect{A}_t=\tr{A\rho(t)}}$ 
of an observable $A$ obeys the differential equation
\begin{equation}
  \label{eq:ME19}
  \frac{d}{dt}\expect{A}_t=
   \frac{1}{i\hbar}\expect{[A,H_{\mathrm{bias}}+H_{\mathrm{dd}}]}_t
   +\expect{\mathcal{L}A}_t\,.
\end{equation}
For observables that commute with $J_z$, which is the case for all operators
related to the signal qubit, only the $\bvec{J}^2$ part of $H_{\mathrm{dd}}$
and the $\tau$-term of $\mathcal{L}$ matter, and then the simpler equation
\begin{equation}
  \label{eq:ME20}
\frac{d}{dt}\expect{A}_t=\frac{i\Omega}{3}\expect{[A,\bvec{J}^2]}_t
+\frac{1}{4\tau}\sum_{k=1}^3\expect{\sigma_{kz}A\sigma_{kz}}_t
-\frac{3}{4\tau}\expect{A}_t
\end{equation}
applies.
In particular we have
\begin{equation}
  \label{eq:ME21}
  \expect{f(J_z)}_t=\expect{f(J_z)}_0
\end{equation}
for all functions of $J_z$. 
The product 
\begin{equation}
  \label{eq:ME22}
\Omega\tau=\frac{\mu_0}{4\pi}\frac{\hbar\Gamma}{a^3(ga)^2}  \propto a^{-5}
\end{equation}
states the relative size of the time constants in Eq.~(\ref{eq:ME20}). 
It depends rather strongly on the distance $a$ between the atoms; 
we have $\Omega\tau\simeq10^{9}$ 
for the values used earlier (${a=883\,\mathrm{nm}}$,
${1/\Gamma=20\,\mathrm{ms}}$, ${ga=9\times10^{-16}\,\mathrm{T}}$).

\subsection{Time dependence of RFF-qubit variables}\label{sec:3D}
By making use of Eq.~(\ref{eq:ME20}), we can now calculate
$\expect{P_{j=1/2}}_t$, the probability of finding the three-atom system in
the ${j=1/2}$ sector at time $t$, and the time-dependent expectation values of
the RFF-qubit Pauli operators of Eqs.~(\ref{Sigma}) and (\ref{eq:Sigma}).
The outcome is
\begin{eqnarray}
  \label{eq:LT3}
  \expect{P_{j=1/2}}_t&=& \frac{1}{3}\Bigl(2+e^{-t/\tau}\Bigr),
\nonumber\\
  \expect{\Sigma_1+i\Sigma_2}_t&=&e^{-\frac{2}{3}t/\tau}
                                 \expect{\Sigma_1+i\Sigma_2}_0\,,
\nonumber\\
  \expect{\Sigma_3}_t&=& e^{-t/\tau}\expect{\Sigma_3}_0\,,
\end{eqnarray}
if the system is initially in the ${j=1/2}$ sector, where the idler and signal
qubits reside.
The expressions for $\expect{P_{j=1/2}}_t$ and $\expect{\Sigma_3}_t$ are exact
solutions of the respective versions of Eq.~(\ref{eq:ME20}), and the result
for $\expect{\Sigma_1+i\Sigma_2}_t$ is an approximation that neglects terms of
relative size $(\Omega\tau)^{-2}\simeq10^{-18}$. 

\begin{figure}
\centerline{\includegraphics{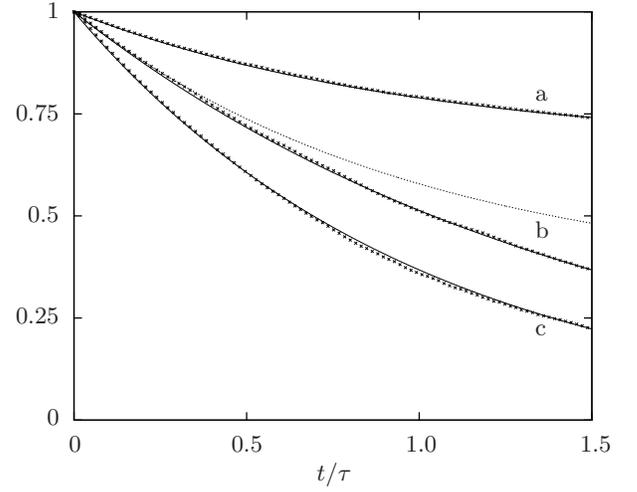}}
\caption{\label{fig:compare}%
Comparison of the data from a numerical simulation 
with the analytical results of Eqs.~(\ref{eq:LT3}).
Curve ``a'' displays $\expect{P_{j=1/2}}_t$, curves ``b'' show
$\expect{\Sigma_1}_t$ for $\expect{\Sigma_1}_0=1$, and curve ``c'' is for
$\expect{\Sigma_3}_t$ with $\expect{\Sigma_3}_0=1$.
The crosses are from a simulation of the dynamics, averaged over 1000 runs.
The solid-line curves represent the analytical results of Eqs.~(\ref{eq:LT3}).
The dotted ``b'' curve shows what one would get for $\expect{\Sigma_1}_t$ if 
$\Omega$ vanished rather than being large on the scale set by $\tau$; we
observe that the inter-atomic dipole-dipole interaction accelerates the decay
of $\expect{\Sigma_1}_t$. 
}
\end{figure}

Supporting evidence is provided by numerical simulations \cite{Lorch:10}. 
These are done by generating a RFF state at initial time $t=0$ and letting the
state evolve with stochastic random noisy magnetic fields on the three atoms. 
All components of the noise are generated randomly at each time step from a
gaussian distribution, and the Maxwell's equations are strictly imposed on the
magnetic fields.  
Figure~\ref{fig:compare} shows both the analytical and the numerical results of
the evolution of the RFF qubit for an arbitrary initial RFF state.
We note that there is very good agreement between the results of the
simulation and the analytical solution of the master equation. 

Quantum information stored in the RFF qubit is degraded substantially only
after a good fraction of $\tau$ has elapsed.
But since $\tau=2\times10^{10}\,\mathrm{s}$ is more than 
$600\,\mathrm{years}$, we conclude that the effect of the inhomogeneous 
magnetic stray fields is of absolutely no concern.
Put differently, the experimenter need not take special measures to suppress
the stray fields.

\subsection{Decoherence of a single-atom qubit}\label{sec:3E}
If --- rather than making good use of the three-atom RFF signal qubit --- one
encoded quantum information into the ${f=1/2}$ ground state of a single $^6$Li
atom, the effect of the random magnetic stray field would be described by the
single-atom master equation
\begin{equation}
  \label{eq:SA1}
  \frac{\partial}{\partial t}\rho(t)=\frac{i\omega_0}{2}[\rho(t),\sigma_z]
           +\frac{1}{4\tau_1}\bigl[\sigma_z,[\rho(t),\sigma_z]\bigr]
\end{equation}
with
\begin{equation}
  \label{eq:SA2}
  \tau_1=\frac{2\tau\tau'}{\tau+\tau'}=\frac{3}{4}\Bigl(\frac{ga}{b}\Bigr)^2\tau
=\Gamma\Bigl(\frac{3\hbar}{2\muB b}\Bigr)^2\ll\tau\,.
\end{equation}
The resulting time-dependent expectation values are
\begin{eqnarray}
  \label{eq:SA3}
  \expect{\sigma_x+i\sigma_y}_t&=&e^{i\omega_0t}e^{-t/\tau_1}
                                  \expect{\sigma_x+i\sigma_y}_0\,,\nonumber\\
  \expect{\sigma_z}_t&=&\expect{\sigma_z}_0\,,
\end{eqnarray}
so that the quantum information can be stored for a fraction of time~$\tau_1$.

The bias field stabilizes the $z$ component: It separates the spin-up and
spin-down states in energy by $\hbar\omega_0$ and so prevents 
transitions between them --- this is, of course, the essence of the
rotating-wave approximation of Eq.~(\ref{eq:ME11}). 
Therefore, one could encode a \emph{classical} bit in a single spin-1/2 atom and
protect it from the stray magnetic field.~\cite{twoatoms}

Without the bias field, the master equation
\begin{equation}
  \label{eq:SA4}
  \frac{\partial}{\partial t}\rho(t)
   =\frac{1}{4\tau_1}\bigl[\bvec{\sigma}\cdot,[\rho(t),\bvec{\sigma}]\bigr]
\end{equation}
applies.
Its solution
\begin{equation}
  \label{eq:SA5}
  \rho(t)=\frac{1}{2}\bigl(1+e^{-2t/\tau_1}
\expect{\bvec{\sigma}}_0\cdot\bvec{\sigma}\bigr)
\end{equation}
shows that the state decays toward the completely mixed state with a life time
of $\tau_1/2$.
It follows that, in addition to preserving the $z$ component, the bias field
also slows down the decay of the $x$ and $y$ components by a factor of two. 

For the example used in Sec.~\ref{sec:3A} --- a fluctuating $0.1\,\mathrm{mA}$
current in a wire at a distance of $50\,\mathrm{cm}$ --- we have
${ga/b\simeq10^{-6}}$ and obtain ${\tau_1\simeq2\times10^{-12}\tau}$. 
Even for the very large value of $\tau$ found above,
${\tau\simeq2\times10^{10}\,\mathrm{s}}$, 
the lifetime of the single-atom qubit is quite
short: ${\tau_1\simeq40\,\mathrm{ms}}$.
Clearly, the well-protected RFF qubit of the three-atom system has an
advantage over the unprotected single-atom qubit:
The stray fields, which are of no concern for the RFF qubit, have a
devastating effect on the single-atom qubit.

More relevant than the lifetime $\tau_1$ is the duration of the initial period
of high fidelity.
We recall that the fidelity of two single-qubit states, specified by their
respective Pauli vectors, is given by
\begin{equation}
  \label{eq:1qb-fidelity}
  F(\rho_1,\rho_2)=\sqrt{\frac{1}{2}(1+\bvec{s}_1\cdot\bvec{s}_2)
+\frac{1}{2}\sqrt{1-\bvec{s}_1^2}\sqrt{1-\bvec{s}_2^2}}\,.
\end{equation}
For the fidelity $F(t)\equiv F\bigl(\rho(t),\rho(0)\bigr)$ between the
initial qubit state and the state at later time $t$, we have the lower bound
\begin{equation}
  \label{eq:1qb-fidelity'}
  F(t)\geq\left\{
    \begin{array}{l@{\mbox{\ for Eq.\ }}l}\displaystyle
      \sqrt{\frac{1}{2}\bigl(1+e^{-t/\tau_1}\cos(\omega_0t)\bigr)} 
      & (\ref{eq:SA1})  \\[2ex]\displaystyle
      \sqrt{\frac{1}{2}\bigl(1+e^{-2t/\tau_1}\bigr)} 
      & (\ref{eq:SA4})
    \end{array}\right.
\end{equation}
so that a fidelity of, say, 0.999 is only guaranteed for a fraction of a
millisecond.
By contrast, the RFF qubit would have a fidelity of 0.9999 or better
for several months if nothing mattered except for the magnetic stray field. 

We note that the ratio of $\tau_1$ and $\tau$ is solely determined by the
comparison of the distance between the atoms and the distance of the atoms
from the source of the noise, for which we have been using $883\,\mathrm{nm}$
and $50\,\mathrm{cm}$, respectively, implying 
$ga/b=900\times10^{-9}/50\times10^{-2}\simeq2\times10^{-6}$.
Therefore, the conclusion that 
${\tau_1/\tau\simeq(2\times10^{-6})^2\simeq10^{-12}}$ holds
irrespective of the actual physical process that generates the stray magnetic
field as long as the noise source is half a meter away.

\section{Dipole-dipole interaction}\label{sec:4}
Two assumptions of ideal geometry enter the derivation of the effective
Hamiltonian for the dipole-dipole interaction in Eq.~(\ref{eq:h}): That the
atoms are located at the corners of a perfect equilateral triangle; and that
the bias field is exactly perpendicular to the plane of the atoms.
Let us now consider the consequences of imperfections on both counts.

\subsection{Center-of-mass probability distribution}\label{sec:4A}
As illustrated by the probability clouds in Fig.~\ref{fig:triangle}, the atoms
do not have definite positions but rather probability distributions for their
centers of mass, given by the ground-state wave functions of the trapping
potentials.
We assume that, for the purpose at hand, 
the respective trapping potentials are reasonably well
approximated by isotropic harmonic oscillator potentials, so that each atom
has a gaussian probability distribution,
\begin{equation}
  \label{eq:CM1}
  p(\bvec{r})=\bigl(\sqrt{2\pi}\,w\bigr)^{-3/2}e^{-\frac{1}{2}r^2/w^2}\,,
\end{equation}
where $\bvec{r}=0$ is the position of the trap center and $w$ is the width of
the gaussian.  
The oscillator frequency $\omega_{\mathrm{trap}}/(2\pi)$ of the trap is
related to $w$ and the mass $M$ of the atom by 
\begin{equation}
  \label{eq:CM2}
  \omega_{\mathrm{trap}}=\frac{\hbar}{2Mw^2}\,,
\end{equation}
which is obtained by fitting the potential around the bottom of the trap to a
harmonic-oscillator potential.
For the potential of Sec.~\ref{sec:6} below, 
$\omega_{\mathrm{trap}}=2\pi\times0.3\,\mathrm{MHz}$, so that the width of the
gaussian wave function is $w=75\,\mathrm{nm}$.
Accordingly, here, earlier in Fig.~\ref{fig:triangle}, and in what follows, 
we take the width $w$ to be about one-twelfth of the distance $a$ between the 
atoms.

When comparing the restoring force of the oscillator potential, 
$M{\omega_{\mathrm{trap}}}^2r$, with the dipole forces exerted by the partner
atoms, $\hbar\Omega/a$, we find that the balance of forces would shift the
equilibrium position by an amount of the order of
\begin{equation}
  \label{eq:CM3}
  \frac{\hbar\Omega}{M{\omega_{\mathrm{trap}}}^2a}
          =\frac{2\Omega}{\omega_{\mathrm{trap}}}\frac{w^2}{a}
          \simeq4\times10^{-9}w\,,
\end{equation}
which is a completely negligible effect. 
We also note that, depending on the joint spin state of the three atoms, 
the shift is in different directions, and the center-of-mass
degrees of freedom get entangled with the spin degrees of freedom but, since 
the shift is such a tiny fraction of the position spread $w$, 
this entanglement is so weak that it can be safely ignored.
As a consequence, the center-of-mass motion is decoupled from the dynamics of
the spins, and probability distributions as in Eq.~(\ref{eq:CM1}) apply to the
atoms at all times.  

The total statistical operator for the three-atom system is then the product 
${\varrho(t)\varrho_{\textsc{cm}}^{\ }}$ of the spin factor $\varrho(t)$ of
Sec.~\ref{sec:3C} and a static center-of-mass factor $\varrho_{\textsc{cm}}^{\ }$.
The von Neumann equation for $\varrho(t)$ is obtained by
tracing over the center-of-mass variables,
\begin{equation}
  \label{eq:CM3'}
  \frac{\partial}{\partial t}\varrho(t)
=\frac{i}{\hbar}\bigl[\varrho(t),
\tr[{\textsc{cm}}]{\varrho_{\textsc{cm}}^{\ }H^{\ }_{\mathrm{tot}}}\bigr]\,,
\end{equation}
where $H^{\ }_{\mathrm{tot}}=H^{\ }_{\textsc{cm}}+H^{\ }_{\mathrm{dd}}+%
H^{\ }_{\mathrm{bias}}+H^{\ }_{\mathrm{noise}}$ is the total Hamiltonian.
Of its four terms, the dipole-dipole interaction energy 
$H^{\ }_{\mathrm{dd}}$ and the noise part $H^{\ }_{\mathrm{noise}}$ involve
both spin variables and center-of-mass variables.
In view of the lesson learned in Sec.~\ref{sec:3}, however, 
there is no need to deal with  $H^{\ }_{\mathrm{noise}}$ in detail. 

We consider the center-of-mass average of the contribution from atoms~1 and
2 to $H_{\mathrm{dd}}$,
\begin{eqnarray}
  \label{eq:CM4}
\tr[{\textsc{cm}}]{\varrho_{\textsc{cm}}^{\ }H^{(12)}_{\mathrm{dd}}}&=&
-\frac{\mu_0}{4\pi}\Bigl(\frac{\muB}{3}\Bigr)^2
\bvec{\sigma}_1\cdot\bvec{\nabla}_{\bvec{a}}\,
\bvec{\sigma}_2\cdot\bvec{\nabla}_{\bvec{a}}
\nonumber\\
&&\times
\int\,(d\bvec{r}_1)(d\bvec{r}_2)\,
\frac{p_1(\bvec{r_1})p_2(\bvec{r}_2)}{|\bvec{a}-\bvec{r}_1+\bvec{r}_2|}\,,
\nonumber\\
\end{eqnarray}
where $\bvec{a}$ is the vector from the trap center for atom~1 to the trap
center for atom~2.
Allowing for different widths of the two gaussians, the integration yields
\begin{equation}
  \label{eq:CM5}
  \int\,(d\bvec{r}_1)(d\bvec{r}_2)\,
\frac{p_1(\bvec{r_1})p_2(\bvec{r}_2)}{|\bvec{a}-\bvec{r}_1+\bvec{r}_2|}=
\frac{1}{a}\erf\Biggl(\frac{a}{\sqrt{2w_1^2+2w_2^2}}\Biggr)
\end{equation}
with the standard error function $\erf(\ )$.
Its asymptotic form
\begin{equation}
  \label{eq:CM6}
  \erf(z)=1-\frac{\ds e^{-z^2}}{\sqrt{\pi}\,z}+\cdots\quad\mbox{for $z\gg1$}
\end{equation}
tells us that the right hand side of Eq.~(\ref{eq:CM5}) differs from $1/a$ by
a term of relative size $10^{-16}$ for $w_1\simeq w_2\simeq a/12$.
It follows that
\begin{equation}
  \label{eq:CM7}
  \tr[{\textsc{cm}}]{\varrho_{\textsc{cm}}^{\ }H^{(12)}_{\mathrm{dd}}}=
\frac{\mu_0}{4\pi}\Bigl(\frac{\muB}{3}\Bigr)^2\bvec{\sigma}_1\cdot
\Biggl(-\bvec{\nabla}_{\bvec{a}}\bvec{\nabla}_{\bvec{a}}\frac{1}{a}\Biggr)
\cdot\bvec{\sigma}_2
\end{equation}
in the present context, which is exactly the $(k,l)=(1,2)$ term 
in Eq.~(\ref{eq:f}), and the center-of-mass probability distribution is of no
further concern.

\subsection{Non-ideal geometry}\label{sec:4B} 
But we need to account for the unavoidable imperfections of any experimental
realization: The triangle formed by the trap centers for the tree atoms is not
exactly equilateral, and the plane of the actual triangle is not exactly
perpendicular to the $z$-axis defined by the bias field.

First, with $a_{kl}$ denoting the distance between the $k$th and the
$l$th trap center, we define the average distance $a$ by means of
\begin{equation}
  \label{eq:NIG1}
  \sum_{(k,l)}\Bigl(\frac{a}{a_{kl}}\Bigr)^3=3\,,
\end{equation}
and measure the deviation of $a_{kl}$ from $a$ by the small parameter
$\alpha_{kl}$,
\begin{equation}
  \label{eq:NIG2}
  \Bigl(\frac{a}{a_{kl}}\Bigr)^3=1-3\alpha_{kl}\,,
  \qquad\sum_{(k,l)}\alpha_{kl}=0\,.
\end{equation}
This average $a$ value is used in Eq.~(\ref{eq:g}) to determine the
dipole-dipole coupling strength $\hbar\Omega$, and we have
\begin{equation}
  \label{eq:NIG3}
  H_{\mathrm{dd}}=\frac{1}{3}\hbar\Omega\sum_{(k,l)}(1-3\alpha_{kl})
            \bvec{\sigma}_k\cdot\bigl(\unitdyad-3\bvec{e}_{kl}\,
                        \bvec{e}_{kl}\bigr) \cdot\bvec{\sigma}_l 
\end{equation}
instead of Eq.~(\ref{eq:f}), where $\unitdyad$ denotes the unit dyadic.
If the atoms are indeed trapped in the minima of an optical potential,
$|\alpha_{kl}|\simeq10^{-2}$ is achievable without resorting to extreme
measures~\cite{David-PC:10}.

Second, nonzero $z$-components of the unit vectors $\bvec{e}_{kl}^{\ }$
require
\begin{equation}
  \label{eq:NIG4}
    \unitdyad-3\bvec{e}_{kl}\,\bvec{e}_{kl}\to
  \frac{1}{2}(3\bvec{e}_z\,\bvec{e}_z-\unitdyad)
\bigl[1-3(\bvec{e}_z\cdot\bvec{e}_{kl})^2\bigr]
\end{equation}
for the step from Eq.~(\ref{eq:f}) to Eq.~(\ref{eq:h}).
Misalignments that exceed $1^\circ$ can be avoided with standard experimental
techniques~\cite{David-PC:10}, so that $|\bvec{e}_z\cdot\bvec{e}_{kl}|=10^{-2}$
is a conservative estimate.

The combined effect of both imperfections is a modification of
$H_{\mathrm{dd}}$, such that
\begin{equation}
  \label{eq:NIG5}
  H_{\mathrm{dd}}=\frac{1}{3}\hbar\Omega\left(3J_z^2-{\bvec{J}}^2\right)
               +\hbar\Omega K
\end{equation}
with
\begin{equation}
  \label{eq:NIG5'}
  K=\sum_{(k,l)}\epsilon_{kl}\frac{1}{2}(\bvec{\sigma}_k\cdot\bvec{\sigma}_l
                         -3\sigma_{kz}\sigma_{lz})\,,
\end{equation}
rather than the $\epsilon_{kl}\equiv0$ version of Eq.~(\ref{eq:h}).
The relative size of the imperfections is measured by 
\begin{equation}
  \label{eq:NIG6}
  \epsilon_{kl}=\alpha_{kl}+(\bvec{e}_z\cdot\bvec{e}_{kl})^2
                  -3\alpha_{kl}(\bvec{e}_z\cdot\bvec{e}_{kl})^2\,,
\end{equation}
wherein, for the values of $|\alpha_{kl}|$ and $|\bvec{e}_z\cdot\bvec{e}_{kl}|$
above, the three terms are of the order $10^{-2}$,
$10^{-4}$, and $10^{-6}$, respectively, and the $\alpha_{kl}$
contribution dominates.

We note in passing that the imperfection parameters $\alpha_{kl}$ and the dot
products $\bvec{e}_z\cdot\bvec{e}_{kl}$ that appear on the right-hand side
of Eq.~(\ref{eq:NIG6}) are not independent of each other. 
Rather, the triangle condition 
$a_{12}\bvec{e}_{12}+a_{23}\bvec{e}_{23}+a_{31}\bvec{e}_{31}=0$ imposes the
restriction 
\begin{equation}
  \label{eq:NIG7}
  \sum_{(k,l)}(1-3\alpha_{kl})^{-1/3}\bvec{e}_z\cdot\bvec{e}_{kl}=0\,,
\end{equation}
where the summation over the pairs is cyclic, that is: $(k,l)=(1,2)$, $(2,3)$,
and $(3,1)$.

The operator $K$ vanishes in the ${j=1/2}$ sector,
\begin{equation}
  \label{eq:NIG8}
  P_{j=1/2}KP_{j=1/2}=0\,,
\end{equation}
and there are no first-order contributions from the $K$-term to the
evolution of the RFF qubit. 
It follows that, during the initial period of high fidelity, the geometrical
imperfections 
contribute in second-order of the small $\epsilon_{kl}$ parameters.

\subsection{Time dependence of RFF-qubit variables}\label{sec:4C}
\begin{figure}
\centerline{\includegraphics{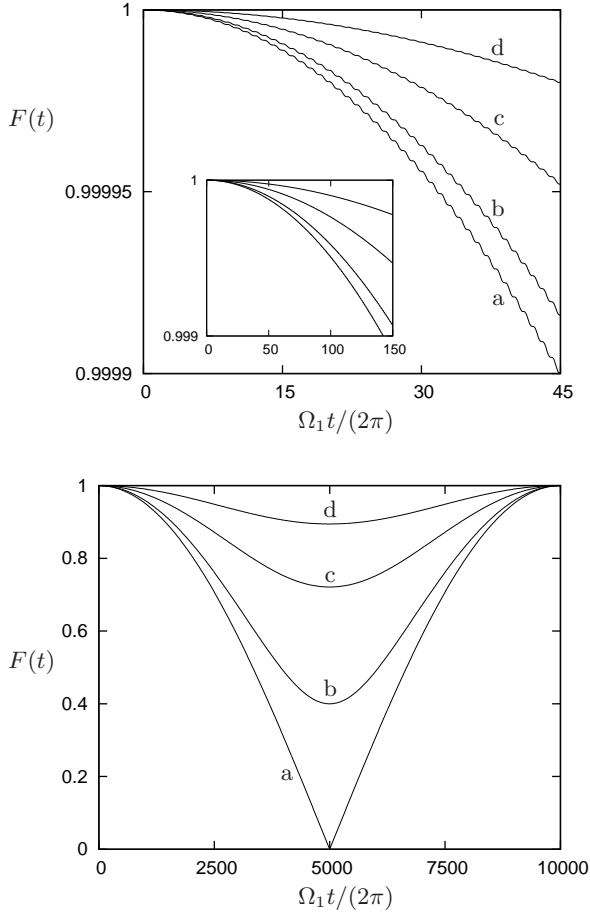}}  
\caption{\label{fig:RFFfidelity}%
Fidelity of the RFF qubit.
For ${\Omega_2/\Omega_1=10^{-4}}$, the plots show $F(t)$ 
of Eq.~(\ref{eq:TD8a}) and its lower bound of Eq.~(\ref{eq:TD8b}) 
for ${t<45\times2\pi/\Omega_1}$ (top plot),  
for ${t<150\times2\pi/\Omega_1}$ (inset in the top plot), and for
${t<2\pi/\Omega_2}$ (bottom plot).
Curve `a' is the lower bound on $F(t)$; 
the other three curves are for 
$\bigl\langle\Sigma_1^{(\varphi)}\bigr\rangle_0=0.4$ and
${s(0)=1}$ (curve `b'), ${s(0)=0.8}$ (curve `c'), and ${s(0)=0.6}$ (curve `d').
One can clearly see the small-amplitude short-period oscillations and the
large-amplitude long-period oscillation. 
For the parameter values used throughout the paper, the respective time ranges
are 2, 7, and 450 hours.}
\end{figure}

For a quantitative analysis, we employ the master equation that results when
Eq.~(\ref{eq:ME18}) is modified in accordance with the observations made here
and above,
\begin{eqnarray}
  \label{eq:TD1}
    \frac{\partial}{\partial t}\rho(t)&=&
      i\omega_0[\rho(t),J_z]
      +i\Omega [\rho(t),{J_z}^2-\tfrac{1}{3}\bvec{J}^2+ K]
\nonumber\\&&\mbox{}
         +\frac{1}{\tau_1}\bigl[J_z,[\rho(t),J_z]\bigr]\,,
\end{eqnarray}
where we put $1/\tau\to0$, $2\tau'\to\tau_1$ in the Lind\-blad operator 
of Eq.~(\ref{eq:ME16}).
While the double-commutator term leads to the fast decay of single-atom spin
coherence, as we saw above in Sec.~\ref{sec:3E}, it is of no consequence for
the RFF qubit because all RFF observables as well as their commutators with
$\bvec{J}^2$ commute with $J_z$. 
It follows further that the expectation value $\expect{A}_t$ of a RFF variable
$A$, with the three-atom system initially prepared in the ${j=1/2}$ sector, is
given by
\begin{equation}\label{eq:TD2}
\expect{A}_t=\expect{{U_{\mathrm{eff}}(t)}^{\dagger}AU_{\mathrm{eff}}(t)}_0  
\end{equation}
with
\begin{eqnarray}
  \label{eq:TD3}
  U_{\mathrm{eff}}(t)&=&P_{j=1/2}\,
              \exp\bigl({-i\Omega t({J_z}^2-\tfrac{1}{3}\bvec{J}^2+K)}\bigr)
                     \,P_{j=1/2}
\nonumber\\ 
  &=&\frac{1+f(t)}{2}P_{j=1/2}
\nonumber\\&&\mbox{}
     -\frac{1-f(t)}{2}\bigl(\Sigma_1\cos\varphi-\Sigma_2\sin\varphi\bigr)\,,
\end{eqnarray}
where 
\begin{eqnarray}
  \label{eq:TD4}
  f(t)&=&
  \frac{\ds\Omega_1e^{-i\Omega_2t}+\Omega_2e^{i\Omega_1t}}{\ds\Omega_1+\Omega_2}\,,
\nonumber\\
\left.\begin{array}{l}\Omega_1 \\ \Omega_2\end{array}\right\}&=&
\frac{\Omega}{2}
\Bigl(\sqrt{(1-\epsilon)^2+8|\kappa|^2}\pm(1-\epsilon)\Bigr)
\end{eqnarray}
for
\begin{eqnarray}
  \label{eq:TD5}
  \epsilon&=&\epsilon_{12}+\epsilon_{23}+\epsilon_{31}>0\,,\nonumber\\
    \kappa&=&\epsilon_{12}+q^2\epsilon_{23}+q\epsilon_{31}=e^{i\varphi}\kappa^*\,.
\end{eqnarray}
For $|\alpha_{kl}|\simeq|\bvec{e}_z\cdot\bvec{e}_{kl}|\simeq10^{-2}$ in
Eq.~(\ref{eq:NIG6}), we have first $\epsilon\simeq|\kappa|^2\simeq10^{-4}$ 
and then $\Omega_1\simeq\Omega$ and 
$\Omega_2=2|\kappa|^2\Omega^2/\Omega_1\simeq10^{-4}\Omega$.
Accordingly, $f(t)$ is the sum of a long-period oscillation with a large
amplitude and a short-period oscillation with a small amplitude; see
Fig.~\ref{fig:RFFfidelity}. 

The resulting time-dependent expectation values of the RFF observables are
\begin{eqnarray}
  \label{eq:TD6}
  \expect{P_{j=1/2}}_t&=&\frac{1+|f(t)|^2}{2}-\frac{1-|f(t)|^2}{2}
  \expect{\Sigma_1^{(\varphi)}}_0\,,
\nonumber\\
  \expect{\Sigma_1^{(\varphi)}}_t
  &=&\frac{1+|f(t)|^2}{2}\expect{\Sigma_1^{(\varphi)}}_0
     -\frac{1-|f(t)|^2}{2}\,,
\nonumber\\
    \expect{\Sigma_2^{(\varphi)}-i\Sigma_3^{\ }}_t&=&
    f(t)\expect{\Sigma_2^{(\varphi)}-i\Sigma_3^{\ }}_0
\end{eqnarray}
with
\begin{eqnarray}
  \label{eq:TD6'}
  \Sigma_1^{(\varphi)}&=&\Sigma_1^{\ }\cos\varphi-\Sigma_2^{\ }\sin\varphi\,,
\nonumber\\
  \Sigma_2^{(\varphi)}&=&\Sigma_2^{\ }\cos\varphi+\Sigma_1^{\ }\sin\varphi\,.
\end{eqnarray}
Equations (\ref{eq:TD6}) contain all information about the state of the signal
qubit in the course of time.
We use them to evaluate the purity of the signal-qubit state at time $t$ and its
fidelity with the initial signal-qubit state.
 
The purity $[1+s(t)^2]/2$ of the signal-qubit state is quantified by the
squared length of its Pauli vector,
\begin{eqnarray}
  \label{eq:TD7}
  s(t)^2&=&
       \frac{\expect{\Sigma_1}_t^2+\expect{\Sigma_2}_t^2+\expect{\Sigma_3}_t^2}
            {\expect{P_{j=1/2}}_t^2}\nonumber\\
&=&1-\frac{|f(t)|^2}{\expect{P_{j=1/2}}_t^2}[1-s(0)^2]\,,
\end{eqnarray}
so that an initially pure signal-qubit state, $s(0)=1$, remains pure.
When the initial state is mixed, ${s(0)<1}$, both ${s(t)>s(0)}$ 
and ${s(t)<s(0)}$ are possible, depending on
the relative size of $|f(t)|$ and $\expect{P_{j=1/2}}_t$.
Specifically, we have
\begin{equation}
  \label{eq:TD7'}
  s(t)\gtrless s(0)
\quad\mbox{if}\quad
\bigl(1+|f(t)|\bigr)\biggl(1+\expect{\Sigma_1^{(\varphi)}}_0\biggr)
\lessgtr 2
\end{equation}
and ${|f(t)|<1}$.

The fidelity $F(t)$ of the signal-qubit state at the
later time $t$ with the initial state is given by
\begin{eqnarray}
  \label{eq:TD8a}
  F(t)^2&=&1-\frac{|1-f(t)|^2}{4\expect{P_{j=1/2}}_t}
           \biggl(1-\expect{\Sigma_1^{(\varphi)}}_0^2\biggr)
\nonumber\\&&\mbox{}
          +\frac{|f(t)|-\mathord{\mathrm{Re}} f(t)}{2\expect{P_{j=1/2}}_t}
           \bigl[1-s(0)^2\bigr]\,.
\end{eqnarray}
It is bounded from below by
\begin{equation}
  \label{eq:TD8b}
  F(t)\geq\sqrt{1-\frac{|1-f(t)|^2}{\bigr(1+|f(t)|\bigr)^2}}\,,
\end{equation}
where the equal sign holds for $f(t)\neq1$ 
if, for example, $s(0)=1$ and 
$\bigl\langle\Sigma_1^{(\varphi)}\bigr\rangle_0=2/\bigl(1+|f(t)|\bigr)-1$.
For $\Omega_2\ll\Omega_1$ and $t\ll2\pi/\Omega_2$, this bound is 
\begin{equation}
  \label{eq:TD8c}
  F(t)\geq\cos\frac{\Omega_2t}{2}
          -\frac{\Omega_2}{2\Omega_1}\sin\frac{\Omega_2t}{2}+\cdots
\end{equation}
where the ellipsis stands for terms of order $(\Omega_2/\Omega_1)^2$.
This two-term approximation serves all practical purposes for
$\Omega_2/\Omega_1\simeq10^{-4}$.
The fidelity is assuredly very high during the early period dominated by the
small-amplitude oscillations with frequency  
$\Omega_1/(2\pi)\simeq\Omega/(2\pi)$: We have ${F=0.9999}$ or better for 45
periods of the fast $\Omega_1$ oscillations, and ${F=0.999}$ or better for 140
periods, when $\Omega_2=10^{-4}\Omega_1$.
These matters are illustrated in Fig.~\ref{fig:RFFfidelity}.

\subsection{Compensating for  triangle distortions}\label{sec:4D}
In Secs.~\ref{sec:4B} and \ref{sec:4C} we regarded the imperfection parameters
$\alpha_{kl}$ and $\bvec{e}_z\cdot\bvec{e}_{kl}$ as resulting from the lack of
perfect control over the apparatus, and their values would not be known with
high precision. 
Suppose, however, that the experimenter has diagnosed the set-up and knows the
actual shape of the triangle quite well while having very precise control over
the direction of the magnetic bias field.
She can then attempt to adjust the bias field such that the three
$\epsilon_{kl}$s of Eq.~(\ref{eq:NIG6}) are equal, with the consequence that
${\kappa=0}$ in Eq.~(\ref{eq:TD5}) and ${f(t)\equiv1}$ in Eq.~(\ref{eq:TD4}).
We do not discuss this matter in further detail and are content with
mentioning that, for small values of the $\alpha_{kl}$s, a bias-field
direction $\bvec{e}_z$ with
\begin{equation}
  \label{eq:NIG-1}
  (\bvec{e}_z\cdot\bvec{e}_{kl})^2
   \simeq \sqrt{\frac{2}{3}\sum_{(k,l)}\alpha^2_{kl}}-\alpha_{kl}
\end{equation}
achieves this, where the approximation neglects terms of second and higher
order in the $\alpha_{kl}$s.
With this compensation for the imperfections in the shape of the triangle
by a judicious tilt of the bias field, the ratio $\Omega_2/\Omega_1$ can
be reduced by much, with a corresponding lengthening of the initial period of
high fidelity.

\section{RFF qubit from four spin-1/2 atoms}\label{sec:5}
An alternative construction of a RFF qubit uses four spin-1/2 atoms and their
two-dimensional subspace with $j=0$. 
A pure state of the RFF qubit is then realized by a pure state of the
four-atom system, and decay results from leakage to the sectors with $j=1$ and
$j=2$, which are nine-dimensional and five-dimensional, respectively. 
By contrast, a pure state of the three--spin-1/2--atom RFF qubit corresponds
to a mixed state of the three-atom system with leakage into the space of the
idler qubit and into the $j=3/2$ sector. 
Clearly, the two constructions of the RFF qubit are substantially different. 

\begin{figure}
\centerline{\includegraphics{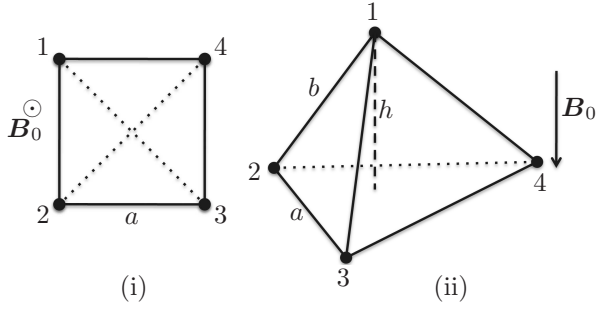}}
\caption{\label{fig:4atoms}%
  RFF qubit constructed from four spin-1/2 atoms.
  (i)Two-dimensional square configuration.
  The dipole-dipole interaction is unavoidably unbalanced here because the
  distance for the two diagonal pairs is larger than the distance for the four
  edge pairs.  
  (ii) Three-dimensional pyramidal configuration. 
  Here, if the height $h$ is chosen such that $b/a=0.661$, the effective
  dipole-dipole interaction has equal strength for all six pairs of atoms.}
\end{figure}

For comparison with the two-dimensional equilateral triangular configuration
of three atoms, let us consider four atoms located at the corners of a
square; see Fig.~\ref{fig:4atoms}(i).
The system is stabilized with a bias magnetic field perpendicular to the plane
of atoms to reduce the decoherence due to the internal pollution from the
dipole-dipole interactions. 
The distance between the two diagonal pairs of atoms is $\sqrt{2}$ times
larger than the distance between the four pairs at the sides. 
Thus, the dipole-dipole interaction is unbalanced between all pairs and it
cannot be made rotationally invariant by the bias magnetic field. 
Explicitly, the effective dipole-dipole interaction is here given by  
\begin{eqnarray}
H_{\mathrm{dd}}&=&\frac{1}{3}\hbar\Omega\biggl(3J_z^2-\bvec{J}^2
-\frac{4-\sqrt{2}}{8}
\bigl[3(\sigma_{1z}\sigma_{3z}+\sigma_{2z}\sigma_{4z})\nonumber\\
&&\hphantom{\frac{1}{3}\hbar\Omega\biggl[}
-(\bvec{\sigma}_1\cdot\bvec{\sigma}_3
+\bvec{\sigma}_2\cdot\bvec{\sigma}_4)\bigr]\biggr)
\end{eqnarray}
with $\hbar\Omega$ as in Eq.~(\ref{eq:g}).
For the same reasons as in the three-atom case, the stray magnetic field is of
no concern, and the system evolves unitarily
\begin{equation}
\rho(t)=e^{-i(H_0+H_{\mathrm{dd}})t/\hbar}\rho_0e^{i(H_0+H_{\mathrm{dd}})t/\hbar}\,.
\end{equation}
The projector onto the $j=0$ subspace of the four-atom RFF qubit is given by 
\begin{eqnarray}
P_{j=0}=\frac{2}{3}(S_{12}S_{34}+S_{13}S_{24}+S_{14}S_{23})\,,
\end{eqnarray}
where the $S_{jk}$s are the singlet state between $j$th and $k$th constituents,
\begin{eqnarray}
S_{jk}=\frac{1}{4}(1-\boldsymbol\sigma_j\cdot\boldsymbol{\sigma}_k)
\end{eqnarray}
for $j,k=1,2,3,4$ and $j\ne k$.

The effective dipole-dipole Hamiltonian has the structure of
Eq.~(\ref{eq:NIG5}), with the operator $K$ now given by
\begin{equation}
  K=\frac{c}{4}\bigl(\bvec{\sigma}_1\cdot\bvec{\sigma}_3
                +\bvec{\sigma}_2\cdot\bvec{\sigma}_4
                -3\sigma_{1z}\sigma_{3z}-3\sigma_{2z}\sigma_{4z}\bigr)\,,
\end{equation}
where $c=(4-\sqrt{2})/6$ accounts for the relative reduction in the strength
of the dipole-dipole interaction for the two diagonal atom pairs. 
Moreover, Eq.~(\ref{eq:TD2}) continues to apply for the expectation value of a
RFF operator $A$, with the four-atom system initially prepared in the $j=0$
sector.
Thereby, the effective unitary evolution operator is now
\begin{eqnarray}
U_{\mathrm{eff}}(t)&=&P_{j=0}e^{-iH_{\mathrm{dd}}t/\hbar}P_{j=0}\nonumber\\
&=&\frac{1+f(t)}{2}P_{j=0}+\frac{1-f(t)}{4}(\Sigma_1'+\sqrt{3}\Sigma_2')
\,,
\quad
\end{eqnarray}
where $\Sigma_1'$, $\Sigma_2'$, and $\Sigma_3'$ are the RFF Pauli operators
for the four-atom system as defined in Eqs.~(25) of Ref.~\cite{Suzuki+2:08},
and $f(t)$ is exactly of the form in Eq.~(\ref{eq:TD4}) with $\Omega_1$ and
$\Omega_2$ replaced by 
\begin{equation}
\left.
  \begin{array}{l}
    \Omega_1\\ \Omega_2
  \end{array}\right\}=
\frac{\Omega}{2}\left(\sqrt{(2-c)^2+8c^2}\pm(2-c)\right)=\left\{
  \begin{array}{r}
    1.78\Omega \\ 0.21\Omega
  \end{array}\right.
\end{equation}
which have a ratio of $\Omega_1/\Omega_2\simeq8.5$, very different from the
ratio of $10^4$ in Eqs.~(\ref{eq:TD4})  and (\ref{eq:TD5}).

\begin{figure}
\centerline{\includegraphics[width=0.95\columnwidth]%
{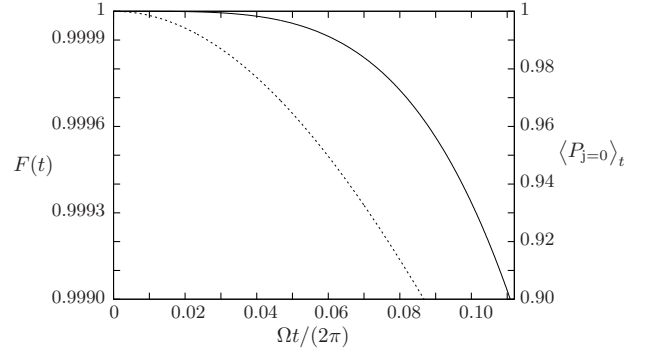}}  
\caption{\label{fig:4atombounds}
Lower bounds for the fidelity $F(t)$ (solid curve) and the expectation value
$\expect{P_{j=0}(t)}$ (dashed curve) for the 
RFF qubit constructed from four spin-1/2 atoms.}
\end{figure}

The expectation values of the RFF operators in the $j=0$ sector can be
obtained analytically, in particular for the expectation value
$\expect{P_{j=0}}_t$ and the RFF qubit fidelity $F(t)$, for which the obvious
analogs of $\expect{P_{j=1/2}}_t$ in Eq.~(\ref{eq:TD6}) and
$F(t)$ in Eq.~(\ref{eq:TD8a}) apply.
With the four-atom version of $f(t)$, the lower bound on $F(t)$ of
Eq.~(\ref{eq:TD8b}) is valid, and we also have $\expect{P_{j=0}}_t\geq|f(t)|^2$. 
Both lower bounds are shown in Fig.~\ref{fig:4atombounds} for the high-fidelity
period of $0\leq\Omega t\leq2\pi\times0.112$. 
We have a fidelity of 0.9999 or better for $t\leq0.062\times2\pi/\Omega$ and
0.999 or better for $t\leq0.11\times2\pi/\Omega$.  

Note, in particular, the substantial probability of losing the four-atom RFF
qubit: After the lapse of $t=0.087\times2\pi/\Omega$, there is a chance of
more than 10\% that the four-atom system has left the $j=0$ sector.
This is a consequence of the rather small $\Omega_1/\Omega_2$ ratio.
By contrast, for the three-atom qubit with $\Omega_1/\Omega_2=10^4$, the
persistence probability $\expect{P_{j=1/2}}_t$ is never less than 0.9996. 

The dipole-dipole coupling strength $\Omega$ is proportional to $1/a^3$, where
$a$ is the length of the sides of the square. 
If we use laser beams with the same wavelengths as used for the three-atom
system in Sec.~\ref{sec:6} to construct the potential for four atoms in a
square geometry, the inter-atomic distance is $a=663\,\mathrm{nm}$ and
Eq.~(\ref{eq:g}) gives $\Omega=2\pi\times16\,\mathrm{mHz}$. 
It follows that we can guarantee $F\geq0.9999$ for about four seconds and
$F\geq0.999$ for about seven seconds. 
This shows that even with a perfect square geometry, the RFF state constructed
from four spin-1/2 atoms decays about 2000 times faster than the qubit
constructed from three spin-1/2 atoms in an imperfect equilateral triangle
configuration. 

The imbalance in the dipole-dipole interaction strength between the
pairs of atoms could be
removed by using the three-di\-mens\-ional pyramidal configuration of
Fig.~\ref{fig:4atoms}(ii). 
To fight against the internal magnetic pollution we apply a bias magnetic
field perpendicular to the plane of three of the atoms which are arranged in
an equilateral triangle, with the fourth atom above the center of the triangle.
By adjusting the height of the pyramid such that the ratio between the atomic
distances is $b/a=0.661$, a rotationally invariant effective dipole-dipole
potential is obtained.  
The lifetime of the RFF qubit for this four-atom geometry is comparable to the
lifetime of the RFF qubit constructed from three atoms in the
equilateral-triangle configuration examined in the previous sections. 
But the practical realization of the peculiar pyramidal arrangement, a distorted
tetrahedron with reduced height, is much more challenging than the equilateral
triangle. 

Clearly, there is no advantage in using the RFF qubits made from four atoms
over the RFF qubits made from three atoms. 
Rather, the simpler three-atom system is preferable.

\section{Structure of the optical lattice}\label{sec:6}
\begin{figure}
\centerline{\includegraphics{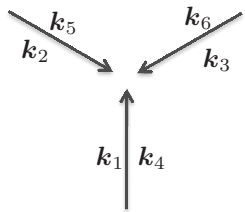}}  
\caption{\label{fig:6lasers}%
Six coplanar laser beams consist of two sets of three coherent beams; 
the angle between beams within each set is $2\pi/3$.
The respective wave vectors have lengths 
$|\bvec{k}_1|=|\bvec{k}_2|=|\bvec{k}_3|=2\pi/\lambda$ and
$|\bvec{k}_4|=|\bvec{k}_5|=|\bvec{k}_6|=2\pi/\lambda'$. 
Different lattice structures can be created by alternating the phases of the
laser beams. 
For the lattice of our design, we keep the set of beams with wavelength
$\lambda$ to be in phase and the phases for the other three beams with
wavelength $\lambda'$ are $2\pi/3$, $0$, and $-2\pi/3$.}
\end{figure}

\begin{figure}
\centerline{\includegraphics{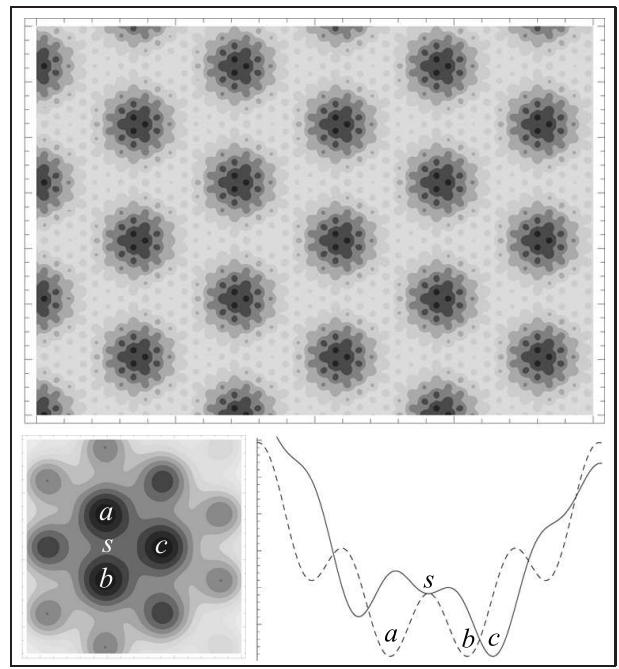}}  
\caption{\label{fig:olatticeCon8}%
The top figure shows the contour plot of the overall optical potential, and
the three global minima at each lattice site of the big triangular lattice
indicate where the trio of atoms can be trapped. 
The bottom left figure gives a more detailed view of the site where one trio
is trapped.
The dashed curve in the bottom right figure is a plot of the potential along
the vertical line cutting through the global minima \textit{a} and \textit{b}; 
and the solid curve is a plot of the potential along the horizontal line
cutting through the other global minima \textit{c}. 
The two cuts intersect at the saddle point \textit{s}.
}
\end{figure}

One kind of optical lattices that could be used is a modification of the
Kagome lattice, where we have an equilateral triangular lattice of which every
site is formed by three spin-1/2 atoms arranged in a small equilateral
triangle. 
A possible physical construction of such a lattice is to use two sets of three
coplanar coherent laser beams and the angle between the beams within each
coherent set is $2\pi/3$. By arranging them in the configuration shown in
Fig.~\ref{fig:6lasers} and adjusting the phases, an optical trapping potential
with the contour plot shown in Fig.~\ref{fig:olatticeCon8} can be produced. For
the potential presented in Fig.~\ref{fig:olatticeCon8}, we chose to keep the
phases of the set of beams with the longer wavelength $\lambda$ to be the same
and the phases for the other three beams with the shorter wavelength
$\lambda'$ are $2\pi/3$, $0$, and $-2\pi/3$, but this choice of phases is not
unique. 

Under the joint consideration of being able to address each RFF qubit
individually,  
of the necessity of high laser intensity to trap the $^6$Li atoms, and of the
requirement of a low probability that the atoms scatter photons from the
trapping lasers, we propose to use the $\textrm{CO}_2$ laser, from which the
desired wavelengths can be generated by frequency doubling (or tripling) in
nonlinear media. 
$\textrm{CO}_2$ lasers with wavelength 10.6$\mu$m are widely used and a high
beam intensity is routinely achieved, and optical trapping of $^6$Li is
reported~\cite{OHara+6:99}. 
We can thus have $\lambda=10.6\,\mu\mathrm{m}$ and 
$\lambda'=\lambda/8=1.33\,\mu\mathrm{m}$ for the wavelengths in 
Fig.~\ref{fig:6lasers}. 
For the produced lattice the atomic distance between atoms within the trio
forming one RFF qubit is $a=883$nm and the distance between the RFF qubits is
$7.1\mu$m. 
The potential strength associated with the low-frequency laser is four times
that of the high-frequency laser for the example of the
potential given in this section.  

The recoil temperature of the $D_1$ (or $D_2$) line of $^6$Li 
is $T_{\mathrm{rec}}=3.5\,\mu\mathrm{K}$. 
To have an estimation of the trapping frequency, the function of the potential
at the minima is fitted with a spherically symmetric harmonic potential. 
With the ratio between the laser intensity and the reduced saturation intensity 
$I_0/I_s=10^8$, the frequency of the harmonic potential is about
$\omega_{\mathrm{trap}}\simeq1.8\,\mathrm{MHz}$. 
The energy separation of the first excited state and the ground state of the
harmonic potential is given by
$\hbar\omega_{\mathrm{trap}}\simeq1.8\times10^{-28}\,\mathrm{J}$ 
($\widehat{=}13\,\mu\mathrm{K}$), which is
about five times the recoil energy. 
The polarizability of $^6$Li is $24.3\times10^{-24}\,\mathrm{cm}^3$, which
yields a scattering rate below $10^{-3}\,\mathrm{s}^{-1}$ at this intensity,
corresponding to a scattering time of more than $1000\,\mathrm{s}$ for one
photon per atom. 
Consequently, the recoil heating is negligible. 
Upon assigning a gaussian profile to the center-of-mass wave function of the
atoms, the spread of the wave function is estimated to be $w=75\,\mathrm{nm}$
for the trapping frequency $\omega_{\mathrm{trap}}$ above.

The depth of the optical trap is roughly $1.1\times10^{-27}\,\mathrm{J}$,
corresponding to a temperature of about $80\,\mu\mathrm{K}$ and more than
twenty times the recoil energy. 
The required intensity of the $10.6\,\mu\mathrm{m}$ $\textrm{CO}_2$ laser is
about $I_0=0.2\,\mathrm{W}/\mu\mathrm{m}^2$ and for the $1.33\,\mu\mathrm{m}$
laser it is about $I_0'=0.03\,\mathrm{W}/\mu\mathrm{m}^2$. 

The numbers given above for the trap are for the sample potential presented
here, which serves the purpose of demonstrating that such a lattice can be
had.
The properties of the trapping potential depend on the laser intensities $I_0$
and $I_0'$. 
Other methods for making an array of atomic trios are conceivable.
This is a hardware issue and all details are determined by the experimental 
set-up at hand.

\section{Summary and discussion}\label{sec:7}
\subsection{Summary} 
We studied the effect of stochastic magnetic stray fields on the RFF qubit 
made from three spin-1/2 atoms and found that the RFF qubit decoheres very
slowly although the spin states of the individual atoms decay very quickly.  
The only coupling of the spin-1/2 atoms to the environment is through their
magnetic dipole moments, so that magnetic stray fields give rise to
uncontrolled changes of the quantum state of the atoms.  
The long lifetime of the RFF qubit results from its insensitivity 
to the over-all magnetic field and its fluctuations because they affect 
all three atoms equally and, therefore, do not affect the RFF qubit at all. 
Decoherence of the RFF qubit originates in spatial variations of the magnetic
field, but they are subject to the constraints imposed by the Maxwell's
equations. 
For parameter values that are typical for experimental situations, we find
that the RFF qubit can maintain a very high fidelity for months --- if the
magnetic stray field is the only source of decoherence.  

We then analyzed the effect of the dipole-dipole interactions among the three
atoms and imperfections in the geometry of the trapped atoms.  
We found that the inter-atomic interactions bring more decoherence to the RFF
qubit than the fluctuating magnetic stray fields, although the dipole-dipole
interaction itself is a unitary process.  
Nevertheless, the RFF qubit states were shown to be very robust within the
parameter regime  and under our assumptions.  
As an example, we showed that the period of high fidelity, say $F=0.9999$,
with respect to the initial state is roughly two hours.

\subsection{Assumptions} 
Let us review the assumptions that we used to derive the result and discuss
their validity. 

We have assumed that the atoms are trapped in the deep optical lattice at very
low temperature so that their center-of-mass motions are negligible. 
One example for such a desired optical lattice, created by standard laser
techniques, was presented in Sec.~\ref{sec:6}.
We estimated, in Sec.~\ref{sec:4A}, the effect of the center-of-mass motions
and found that it is much smaller (i.e., $10^{-16}$) than the effect of the
dipole-dipole interactions for the optical lattice considered.  

The most challenging element seems to be to maintain the stable lasers for the
optical lattice in order to observe the long-time evolution of the RFF qubit. 
In a real experiment the collisions with rest-gas atoms are also inevitable
and they may very well limit the lifetime of the RFF qubit in practice.
We note that drifts of the lasers in time would not spoil the long lifetime 
as long as all lasers are locked in phase. 
This is because the time scale of these parameter changes is much slower and
hence all atoms follow the optical lattice adiabatically.  

The noise model employed in this study is divided into two types. 
One is a fluctuating magnetic stray field arising from unavoidable
imperfections in the surrounding apparatuses such as the Helmholtz coils,
electric wires, and so on. 
The Helmholtz coils used to generate the homogenous bias magnetic field are
identified as the major source for the noise of this kind. 
Other possible fluctuating magnetic fields are much smaller than this and less
inhomogeneous as the respective sources are farther away. 
We then linearized these fluctuating fields around the homogeneous bias field
to analyze the decoherence for the RFF qubit in Secs.~\ref{sec:3A} and
\ref{sec:3C}.  

The other type of noise is due the magnetic dipole-dipole interaction among
the three atoms. 
We have shown that these inter-atomic interactions are a major source of
decoherence for the RFF qubit when imperfections of the experimental set-up
are taken into account. 
In Secs.~\ref{sec:4B} and \ref{sec:4C},
we accounted for deviations from the ideal equilateral triangle configuration
of the three atoms as well as a misalignment of the magnetic bias field and 
found that such insufficiencies still allow for a very 
long lifetime of the RFF qubit.  
In fact, we observed that imperfections in the geometry of the three atoms
could be compensated for by adjusting the direction of the bias magnetic
field, provided that the experimenter has sufficient control over the relevant
parameters.

We should not forget to mention that the rotating-wave approximation was
used, for example, when analyzing the effect of fluctuating magnetic stray
fields.  
This approximation is valid within the energy scale of our set-up, where a
probability for a non-resonant transition is very small. 
The numerical study without the rotating-wave approximation 
also supports the validity of our master-equation analysis. 

All these results are, of course, derived with the assumption that 
the parameters used in this study are well controlled with a certain precision. 
We however took rather conservative numbers for these parameters 
so that we can estimate a realistic lifetime of the RFF qubit. 
This also leaves some room for improving the lifetime of the RFF qubit in
further studies.

\subsection{Alternatives}
In the scheme presented here, the RFF state is made from three spin-1/2 atoms. 
But this is not the only way to construct RFF qubits. 
We can equally well use three identical atoms of any non-zero ground-state
spin $j$, where the RFF signal qubit lives in the subspace of total angular
momentum $3j-1$, which has two states for each $m$ value.  
The idler space is then $(6j-1)$-dimensional. 
Analogously, RFF qutrits can be constructed in the sector of total angular
momentum $4j-1$ from four identical atoms with ground state spin $j$.  
Such alternative constructions offer considerable flexibility in choosing the
isotope for a practical implementation. 

As we mentioned in Sec.~\ref{sec:5}, other geometries and four atoms are not
better than the case of three atoms in the equilateral triangle, but they are
still good for the purpose of storing quantum information.  
For example, three or four atoms on a line, with the bias field in the right
direction, could be an easier choice for a trapped-ion experiment.  
Likewise, physical systems other than cold atoms should in principle give 
a long lifetime if information is encoded in the RFF subsystem.   
There are advantages and disadvantages depending upon the physical system one
chooses.  
All these are largely unexplored territories that need further surveying.

\subsection{Outlook} 
Having thus shown how to construct robust units for storing quantum
information, the problems of how to encode and read out the information, 
and how to process it in such a set-up, need to be addressed. 
We will report progress on this front in due course.

\begin{acknowledgments}
Centre for Quantum Technologies (CQT) is a Research Centre of Excellence
funded by Ministry of Education and National Research Foundation of
Singapore. 
N.L. would like to thank the Studienstiftung des deutschen Volkes for
financial support. 
J.S. is supported by NICT and MEXT. 
Both N.L. and J.S. would like to thank CQT for the kind hospitality. 
H.R. wishes to thank Kae Nemoto and the National Institute of Informatics for
their warm hospitality. 
Both H.R. and B.-G.E. thank Hans Briegel and the Institute for Quantum Optics
and Quantum Information for their warm hospitality.
We gratefully acknowledge Ng Hui Khoon's enlightening comments and friendly
discussions. 
\end{acknowledgments}


\begin{thebibliography}{99}
\bibitem{err.Suzuki+2:08}
The correspondence with the conventions of Ref.~\cite{Suzuki+2:08} is the
following: What is $q$ here, is $\omega_3$ there; 
the labels ${\lambda=0,1}$ here are ${\lambda=1,2}$ there; 
and the operators $Q_0$ and $Q_1$ here are $\Omega_-(1)$ and 
$\Omega_-(2)$ there.
For the record we note that an inadvertent interchange between
${\lambda=1}$ and ${\lambda=2}$ happened in the transition from Sec.~III to
Sec.~IV\,A in \cite{Suzuki+2:08}.
\bibitem{ContactTerm}
There is no need, here or later in Sec.~\ref{sec:4A}, to include the contact
term $\propto\bvec{s}_k\cdot\bvec{s}_l\delta(\bvec{r}_{kl})$.
\bibitem{twoatoms}
When using two atoms, the bias field separates the $\uparrow\uparrow$ and
$\downarrow\downarrow$ states from the $\uparrow\downarrow$ and
$\downarrow\uparrow$ states, which have approximately the same energy 
and can be used for the storage of a qubit. 
A life time of several seconds was achieved in an ion-trap
experiment~\cite{Langer+15:05}.
\end{thebibliography}
\end{document}